\documentclass[oupdraft]{bio}
\usepackage{url}
\usepackage[utf8]{inputenc}
\usepackage{diagbox}
\usepackage{multirow}
\usepackage{graphicx}
\usepackage{setspace}
\usepackage{amsmath}
\usepackage{longtable}
\usepackage[usenames,dvipsnames]{color}
\usepackage{lscape}
\usepackage[toc,page]{appendix}
\usepackage{amssymb}
\usepackage{lineno}
\usepackage{fancyhdr}
\usepackage{lastpage}
\usepackage{arydshln}
\usepackage{soul}
\usepackage{floatrow}
\usepackage{pdflscape}
\usepackage{chapterbib}  
\usepackage{arydshln}
\usepackage{mathptmx}
\usepackage{makecell}
\usepackage{tablefootnote}

\newcommand{\PM}{PM$_{2.5}$ }

\newcommand{\bnmf}{BN$^{2}$MF }
\newcommand{\bnmfc}{BN$^{2}$MF}

\begin{document}

\title{Bayesian non-parametric non-negative matrix factorization for pattern identification in environmental mixtures}

\author{ELIZABETH A. GIBSON$^1,\ast$, SEBASTIAN T. ROWLAND$^1$, JEFF GOLDSMITH$^2$, JOHN PAISLEY$^3$, JULIE B. HERBSTMAN$^1$, MARIANTHI-ANNA KIOURMOURTZOGLOU$^{1}$\\[10pt]
\textit{$^1$Department of Environmental Health Sciences, Columbia University Mailman School of Public Health, New York, NY \\[10pt]
$^2$Department of Biostatistics, Columbia University Mailman School of Public Health, New York, NY \\
$^3$Department of Electrical Engineering, Columbia University Data Science Institute, New York, NY}
\\
{e.a.gibson@columbia.edu}
}

\markboth%
{E. A. Gibson and others}
{\bnmf for pattern identification in environmental mixtures}

\maketitle

\footnotetext{To whom correspondence should be addressed.}

\begin{abstract}
{Environmental health researchers may aim to identify exposure patterns that represent sources, product use, or behaviors that give rise to mixtures of potentially harmful environmental chemical exposures. We present Bayesian non-parametric non-negative matrix factorization (\bnmfc) as a novel method to identify patterns of chemical exposures when the number of patterns is not known \textit{a priori}. We placed non-negative continuous priors on pattern loadings and individual scores to enhance interpretability and used a clever non-parametric sparse prior to estimate the pattern number. We further derived variational confidence intervals around estimates; this is a critical development because it quantifies the model's confidence in estimated patterns. These unique features contrast with existing pattern recognition methods employed in this field which are limited by user-specified pattern number, lack of interpretability of patterns in terms of human understanding, and lack of uncertainty quantification.

To test this method, we generated simulations with increasing levels of complexity in pattern structure and increasingly noisy scenarios. We assessed \bnmfc's coverage of true simulated scores and compared our model's performance with methods frequently used in environmental health. \bnmf estimated the true number of patterns for 99\% of simulated datasets. \bnmfc's variational confidence intervals achieved at least 95\% coverage across all levels of structural complexity with up to 40\% added noise. \bnmf performed comparably with frequentist methods in terms of overall prediction and estimation of underlying loadings and scores. After validation, we applied \bnmf to a mixture of 17 endocrine disrupting chemicals (EDCs) measured in 343 pregnant women in the Columbia Center for Children’s Environmental Health's Mothers and Newborns Cohort.  
}
{Pattern recognition, Bayesian analysis, non-negative matrix factorization, environmental epidemiology, endocrine disrupting chemicals}
\end{abstract}

\section{Introduction}
\label{intro}
Individuals are ubiquitously exposed to multiple classes of chemicals through environmental pollution and consumer products. Despite evidence from bio-monitoring studies indicating that exposures are often highly-correlated \citep{woodruff2011env}, environmental health research has historically examined single chemicals and classes in isolation. The combination of exposures, however, likely exhibits different relationships with potential health outcomes \citep{braun2016can, taylor16}. In recent years, there have been concerns regarding the health effects of environmental mixtures such as air pollution \citep{kioumourtzoglou2013effect}, heavy metals and metalloids \citep{wasserman2018cross}, and endocrine disrupting chemicals (EDCs) \citep{tanner2020early}. Due to common environmental sources or behaviors leading to exposure, potentially shared biological pathways, and similar toxicological effects, environmental health researchers often desire a tool for exposure pattern recognition \citep{gibson2019complex}.

To identify patterns in high-dimensional mixtures, researchers must overcome several challenges. First, the true number of patterns in a given population is not known. Existing methods employed in environmental health research use \textit{a priori} criteria specification (e.g., percent of variance explained by components in principal component analysis (PCA) or lowest Akaike information criterion (AIC) for Non-negative Matrix Factorization (NMF)) to choose the number of components retained or factors selected. This practice opens the door to subjectivity and interpretation, as two researchers using the same method on the same dataset may choose different criteria and, thus, reach different conclusions. Second, chemical concentrations are non-negative and, therefore, cannot be intuitively understood as a negative feature. Methods that return orthogonal solutions, such as PCA, and/or solutions that contain negative values (e.g., PCA, factor analysis, and various matrix factorizations), therefore, do not reflect the process that generates observations, making the solution less interpretable. Potentially negative chemical loadings cannot be easily interpreted as either present or absent, and potentially negative individual scores do not intuitively convey exposure level. Third, pattern recognition in environmental health is often the first step of a two-stage approach to estimate associations between identified patterns and health outcomes. This makes the estimation of uncertainty in the pattern recognition step essential for subsequent construction of appropriate confidence intervals of health effect estimates. Common methods such as PCA, NMF, and factor analysis do not quantify uncertainty in estimation.

In this paper, we introduce Bayesian non-parametric non-negative matrix factorization (\bnmfc) as an approach for identifying patterns in environmental mixtures. \bnmf decomposes observed exposure data (e.g., chemicals across participants in a cohort study or across days in a time series) into a matrix of chemical loadings and a matrix of individual scores on \bnmfc-identified patterns that are much smaller in number than the total number of chemicals. Because non-negativity is an informative feature of environmental exposure data, we enforced it on pattern loadings and individual scores with strictly non-negative priors, making them interpretable on an additive scale with a parts-based representation \citep{lee1999learning}.

Previous work on NMF comes from machine learning. \citet{lee1999learning} first introduced NMF as a method to learn a parts-based representation of data. In text analysis, NMF methods have been primarily applied to identify semantic features, or topics, of words shared across documents \citep{blei2003latent, paisley2014bayesian}. In the context of image recognition and compression, NMF methods have been applied to learn factors that correspond to parts of an original image and to reconstruct images that have been corrupted \citep{sandler2011nonnegative, cemgil2008bayesian}. Within environmental health, the majority of research employing pattern recognition methods has been in air pollution source apportionment, such as the U.S. Environmental Protection Agency's positive matrix factorization (PMF) \citep{paatero94}. The majority of NMF algorithms require the number of factors to be specified by the user. \bnmf stems directly from \citeauthor{holtzman2018machine}' implementation of a Bayesian non-parametric matrix factorization model to determine a suitable rank of factorization from the data. In their work, they employed NMF to identify frequency regions that co-occur across spectrograms to better identify changes in faulting processes leading to earthquakes \citep{holtzman2018machine}.

Our work provides several contributions to existing environmental exposure pattern recognition approaches. To our knowledge, this is the first Bayesian unsupervised statistical method considered for pattern recognition in mixtures of environmental exposures. Previous work in this area has utilized frequentist statistical and machine learning approaches in which the number of patterns in the mixture was specified by the researcher or treated as a tuning parameter. Here, we employed an empirical sparse prior on the number of patterns in the mixture to estimate it from the data. Additionally, we derived variational confidence intervals surrounding individual scores that provide previously unattainable uncertainty quantification. We conducted simulation studies (Section \ref{methods_sim}) in which we compared our method to other pattern identification methods, namely PCA, factor analysis, and two frequentist NMF models with different objective functions. Finally, we applied \bnmf to an environmental mixture of potential EDCs measured in pregnant women in a mother and child cohort from New York City (Section \ref{results_app}). This application highlights \bnmfc's ability to identify the number of underlying patterns in a chemical mixture without guidance from the researcher and yield interpretable results.

\section{Methods}
\label{methods}
In traditional NMF, given an $N \times P$ non-negative matrix $X =\lbrace x_{i j}\rbrace $, where $i = 1:N$ and $j = 1:P$, we seek non-negative matrices $W$ and $H$ such that,
\begin{equation}
\label{nmf_trad}
X_{i j} \approx \sum^K_{k=1} W_{i k} H_{k j},
\end{equation}

\noindent where $K$ is the rank of the solution matrices \citep{lee1999learning}. We will refer to the $N \times K$ matrix $W$ as the \textit{coefficient matrix} of individual pattern scores and the $K \times P$ matrix $H$ as the \textit{dictionary matrix} of chemical loadings on patterns. In the context of environmental mixtures, $X$ typically characterizes a high-dimensional exposure space, where each observation $\mathbf{x}_{i}=\left(x_{i 1}, \ldots, x_{i P}\right)^{\mathrm{T}}$ is a row vector of $P$ exposure variables (e.g., \PM components or EDCs). In the coefficient matrix, the row vector $\mathbf{w}_{i}=\left(w_{i 1}, \ldots, w_{i K}\right)^{\mathrm{T}}$ contains a dimension-reduced representation for the $i^{th}$ observation, where each observation  shares the same set of $K$ patterns, with a unique mixture of pattern membership \citep{wang2012nonnegative}. In the dictionary matrix, each pattern $\mathbf{h}_{k}=\left(h_{k 1}, \ldots, x_{k P}\right)^{\mathrm{T}}$ is a row vector with a unique mixture of chemical loadings.

\subsection{Bayesian Formulation}
\label{methods_bayesian}
Traditional NMF minimizes a squared error objective or an information divergence objective through a multiplicative algorithm \citep{lee1999learning}. The negative divergence penalty is equivalent to the maximum likelihood method in a Poisson model \citep{cemgil2008bayesian,virtanen2008bayesian}.
Given this probabilistic interpretation, \citet{cemgil2008bayesian} described a hierarchical structure with Gamma priors on both coefficient and dictionary matrices, with ${W \sim \operatorname{Gamma}(\alpha_{W}, \beta_{H})}$ and ${H \sim \operatorname{Gamma}(\alpha_{H}, \beta_{H})}$, where each component is drawn independently. This allows a fully Bayesian treatment of the model.


Gamma distributions were originally chosen out of computational convenience because Gamma is the conjugate prior to the Poisson likelihood \citep{cemgil2008bayesian}. In our application, however, the motivation is reversed: Gamma priors fit our data generating process and we accepted the Poisson likelihood as a computational convenience. Measured pollutant exposure concentrations are non-negative, continuous values, as are the conceived coefficient and dictionary matrices. Thus, Gamma priors are well-suited for our latent factors. In practice, applications using NMF with the divergence objective function perform well on non-negative continuous data \citep{brunet2004metagenes, fevotte2009nonnegative}.

\subsubsection{Non-parametric estimation of $k$.}
\label{methods_k}
Following \citet{holtzman2018machine}, we employed a Bayesian non-parametric approach to determine a suitable factorization of the data. We modeled the exposure matrix using a $K$-rank factorization by including $\mathbf{a}$ as an empirical prior,
\begin{equation}
{X \sim \operatorname{Pois}\left(W \operatorname{diag}(\mathbf{a}) H\right)}
\end{equation}
\vspace{-5ex}
\begin{equation*}
{W \sim \operatorname{Gamma}(\alpha_W = \mathbf{1}, \beta_W = \mathbf{1})}, \hspace{1ex}
{\mathbf{a} \sim \operatorname{Gamma}(\alpha_\mathbf{a} = \frac{\mathbf{1}}{K}, \beta_\mathbf{a} = 1)}, \hspace{1ex} 
{H \sim \operatorname{Gamma}(\alpha_H = \mathbf{1}, \beta_H = \mathbf{1})}
\end{equation*}
\vspace{-5ex}
\begin{equation}
    \underbrace{\mathbb{P}(W, \operatorname{diag}(\mathbf{a}), H)}_{\text{Posterior}}
    \propto \underbrace{\operatorname{Pois}(W\operatorname{diag}(\mathbf{a})H)}_{\text{Likelihood}}
    \times \underbrace{\operatorname{Gamma}(\alpha_W, \beta_W) 
    \times \operatorname{Gamma}(\alpha_\mathbf{a}, \beta_\mathbf{a}) 
    \times \operatorname{Gamma}(\alpha_H, \beta_H)}_{\text{Priors}}
\end{equation} 

\noindent Here, $W$ and $H$ begin with the same dimensions as the original implementation (equation~\ref{nmf_trad}), and we introduce $\mathbf{a}$ as a $K$-dimensional vector with a sparse Gamma prior ($\alpha_\mathbf{a} < 1$). $K$ is initialized to equal the mixture size $P$ and is updated iteratively, i.e., it is allowed to vary, and therefore so is the prior for $\mathbf{a}$. The diagonal matrix $\operatorname{diag}(\mathbf{a})$ shrinks out unnecessary rows of $H$ by setting them to zero, or equivalently, it shrinks the corresponding columns of $W$. This non-parametric prior empirically infers the number of factors, $K$, from the data ($K < P$). With the inclusion of $\mathbf{a}$, we will view the $N \times K$ matrix product $W\operatorname{diag}(\mathbf{a})$ as the coefficient matrix and $\mathbf{w}_{i} \circ \mathbf{a}$ as the dimension-reduced representation for the $i^{th}$ observation, where ``$\circ$'' represents element-wise multiplication. We took $\mathbb{E}_{q}\left[W_{i k}\mathbf{a}_{k}\right]$ and $\mathbb{E}_{q}\left[H_{k j}\right]$ as individual scores and chemical loadings, respectively.

\subsubsection{Variational inference.}
\label{methods_vi}
We used variational inference to approximate the latent parameters and their distributions. Variational inference is a fundamental machine learning technique that converts Bayesian posterior inference into an optimization problem \citep{jordan2004graphical,ormerod2010explaining}. It begins at an initial setting of specified independent variational parameters, then optimizes them to find the members of their families that are closest to the exact posterior distributions \citep{wainwright2008graphical}. We define the approximating distributions and the variational objective function in Supplemental Section \ref{suppVI}.

\subsubsection{Variational confidence intervals.}
\label{methods_vci}
We empirically computed 95\% variational confidence intervals around the expected values in the coefficient matrix of individual scores. We first generated 1,000 random draws from the solution distributions $\operatorname{Gamma}(\hat{W}_{1_{i k}}, \hat{W}_{2_{i k}})$, $\operatorname{Gamma}(\hat{\mathbf{a}}_{1_k}, \hat{\mathbf{a}}_{2_k})$, and $\operatorname{Gamma}(\hat{H}_{1_{k j}}, \hat{H}_{2_{k j}})$. We then $\ell_1$-normalized the patterns in the dictionary matrix ($\mathbf{h}_{k \cdot}$) and scaled the patterns in the coefficient matrix ($W\operatorname{diag}(\mathbf{a})_{\cdot k}$) by the corresponding normalization constant. This put all individual scores across patterns on the same scale. We defined the 95\% variational confidence intervals as the 2.5\textsuperscript{th} and 97.5\textsuperscript{th} quantiles of the of the empirical distribution of the scaled scores.

The variance of a parameter obtained through variational inference is known to be narrower than the variance of the true posterior distribution \citep{svensen2005robust}. For this reason, we created bootstrapped confidence intervals to compare with the variational confidence intervals, further described in Supplemental Section~\ref{supp_boot}.

\subsection{Simulation strategy}
\label{methods_sim}
We evaluated the ability of \bnmf to identify the true patterns underlying high-dimensional mixtures, and compared our model's performance to three commonly used frequentist methods. We applied \bnmfc, PCA, factor analysis, and NMF to simulated datasets with known data generating processes designed to reflect realistic underlying patterns in environmental mixtures with increasing complexity.

Generally, the notion of NMF is not identifiable, meaning that the factorization is not unique (up to scaling and permutation) \citep{laurberg2008theorems}. Previous research has introduced separability, a fairly mild assumption about the underlying pattern structure, under which NMF may recover the truth \citep{donoho2004does}. In the context of environmental data, the exposure matrix $X$ is separable if, for each pattern $k$ in the dictionary matrix, $H$, there is some chemical $j$ such that $H_{k j} > 0$ and $H_{k^{\prime} j}=0$ for $k^{\prime} \neq k$ \citep{arora2012learning}. In other words, there is at least one chemical that loads solely on each pattern. This usually holds in real-life data. In the following simulations, this assumption does not always hold, as increased noise prevents separability and some simulations are structurally inseparable (i.e., all chemicals load on more than one pattern). 

\subsubsection{Primary data generation.} 
We generated 12,100 datasets of 1,000 observations each, $\left\{\mathbf{x}_{i}\right\}_{i=1}^{1,000},$ where $\mathbf{x}_{i} = \left(x_{i,1}, \ldots, x_{i,40}\right)^{\mathrm{T}}$ presents an exposure profile with 40 mixture components. We specified four underlying patterns. We created the simulated data from matrix products of $1,000 \times 4$ coefficient matrices and $4 \times 40$ dictionary matrices with added Gaussian noise.

We first generated dictionary matrices with patterns ranging from distinct to completely overlapping. Each chemical was drawn from a Dirichlet distribution so that its loadings summed to one over all patterns. For distinct patterns, ten chemicals had loadings of 1 for each pattern, with no chemical loading on multiple patterns. For overlapping patterns, ten chemicals had `high' loading on one pattern, `medium' on a second pattern, and no loading on the remaining two patterns. These were generated so that the `high' loading would be approximately twice the `medium' loading from one of four distributions: $\operatorname{Dir}(\alpha_1=10, \alpha_2=5, \alpha_3=0, \alpha_4=0)$, $\operatorname{Dir}(\alpha_1=0, \alpha_2=10, \alpha_3=5, \alpha_4=0)$, $\operatorname{Dir}(\alpha_1=0, \alpha_2=0, \alpha_3=10, \alpha_4=5)$, $\operatorname{Dir}(\alpha_1=5, \alpha_2=0, \alpha_3=0, \alpha_4=10)$. `Distinct' and `overlapping' represent the extremes of the simulation process; overlapping patterns are not separable. Nine more datasets were generated by progressively moving four chemicals (one from each pattern) from the distinct to overlapping data generating process.

We next generated coefficient matrices. We drew scores independently from $\operatorname{Lognormal}(\mu = 0, \sigma^{2} = 1)$. We took the product of coefficient and dictionary matrices to generate chemical exposure matrices. We added noise from a Gaussian $\mathcal{N}(\mu = 0, \sigma = s)$, replacing negative values in the sum of exposure and error matrices with zero. The noise level $s$ was defined as a proportion of the `true' standard deviation in the simulation before noise was added. This proportion ranged from 0 to 1 in increments of 0.1. Zero corresponds to a scenario with no noise in measurement, which is not meaningful in an applied setting, as all environmental measurements have noise \citep{van2020reflection}; however, it is useful as a baseline against which to compare results.

Before adding noise, all simulated datasets were of rank 4; after adding noise, they became full rank. Separable simulations were no longer fully separable when noise was added. We generated 100 datasets each for all possible combinations of 11 dictionary matrix structures and 11 noise levels (121 data generating processes), so that all simulations were analyzed at increasing noise levels. For the purpose of this work, we present results for two pattern structures (distinct and overlapping) and three noise levels, 0.2, 0.5, and 1 (600 datasets) in tables and figures and refer the reader to the supplemental materials for the remaining results. The bootstrapped models from Section~\ref{methods_vci} were conducted on the same subset.

\subsubsection{Secondary data generation.} 
To assess the dependence of our results on dataset and pattern dimension, we generated datasets with varying dimensions and specifications. We considered three sample sizes ($N$ = 200, 1,000, 10,000), three feature spaces ($P$ = 20, 40, 100), and three pattern numbers ($K$ = 1, 4, 10). These simulations were compared with the main subset of distinct and overlapping patterns at three noise levels (0.2, 0.5, and 1.0).

\subsubsection{Metrics and measures of comparison.}
\label{methods_metrics}
We ran \bnmfc, PCA, NMF, and factor analysis on all simulated datasets. We ran two versions of NMF, one with the information divergence penalty (NMF-P for Poisson) and one with the squared error objective (NMF-$\ell_2$). For PCA, NMF, and factor analysis, we specified \textit{a priori} criteria for component retention or factor selection. For PCA, we retained the first $n$ components that explained $\geq 80\%$ of the variance in the data, as commonly done in environmental mixtures applications (e.g., \cite{gibson2019overview}). For NMF and factor analysis, we ran three models with varying rank ($K$ = 3, 4, and 5 for the primary simulations) and chose the model with the lowest Bayesian information criterion (BIC). We then compared solutions across methods using relative predictive error, cosine distance \citep{tan2016introduction}, and symmetric subspace distance \citep{wang2006subspace}, further detailed in Supplemental Section~\ref{supp_metrics}.

To assess the validity of the variational confidence intervals, we defined coverage as the proportion of simulated individual scores that fell within their lower and upper bounds. We employed the same scaling steps---$\ell_1$-normalization of the dictionary matrix and corresponding scaling of the coefficient matrix---on the simulations to put them on the same scale as the solutions.

\section{Results}
\label{results}
\subsection{Simulation findings}
\label{results_sim}
\subsubsection{Primary simulations.}
\bnmf chose a $K = 4$ solution for 99\% of the simulations, which is consistent with the simulation design. Of the 137 datasets for which \bnmf was incorrect, 129 of them were simulated with high proportions of added noise (0.9 (n = 20) or 1.0 (n = 109)) relative to the true values. Factor analysis selected the 4 factor model based on BIC at all noise levels, but when no noise was added, factor analysis failed to converge. NMF (both with the $\ell_2$ and divergence penalties) selected the four factor model based on BIC $>$99\% of the time. Like \bnmf, NMF failed more often when noise was high. PCA only retained four components in 34\% of the models. It never chose 4 patterns when the noise proportion was $>$ 0.6.

We assessed model accuracy by comparing predicted values from each model with the underlying truth before noise was added. Mean relative prediction error and standard deviation are shown in Table~\ref{tab_l2}. \bnmf and frequentist NMF outperformed PCA and factor analysis at reconstructing the true data. Factor analysis had the poorest performance overall, with relative prediction error consistently higher than the other methods. PCA performed well when the underlying patterns were orthogonal and there was little noise but not when patterns overlapped or more noise was added. Frequentist NMF performed just as well or slightly better than \bnmf in terms of predicted accuracy; this is not surprising, since more accurate prediction is not the key benefit of the Bayesian approach.

To assess estimation accuracy for loadings and scores, we used relative prediction error for solutions that correctly identified the number of underlying patterns (see Table~\ref{tab_l2}). On this metric, \bnmf and factor analysis outperformed PCA and frequentist NMF at estimating scores, and factor analysis and NMF-P better estimated loadings. PCA performed well on loadings but worse on scores, and often provided a result that was not directly comparable with the truth.

To assess estimation accuracy for loadings and scores in terms of orientation, we used cosine distance for solutions that correctly identified the number of underlying patterns. Mean cosine distance and standard deviation are shown in Table~\ref{tab_cos}. Non-negative methods, both \bnmf and frequentist NMF, generally outperformed PCA and factor analysis at estimating the direction of true patterns. 

Finally, we compared the linear subspaces spanned by estimated scores and estimated loadings with true underlying scores and loadings using the symmetric subspace distance. This metric allowed comparison of results that incorrectly estimated the number of patterns. Mean symmetric subspace distance and standard deviation are shown in Table~\ref{tab_ssd}. On this metric, the non-negative methods again outperformed PCA and factor analysis. Results for secondary simulations varying $N$, $P$, and $K$ are included in Supplemental Section~\ref{sec_sim}.

\subsubsection{Variational confidence intervals.}
Figure~\ref{heatmap} displays median coverage of variational confidence intervals for simulation scenarios taken across entries in 100 datasets per grid square, for all pattern structures and added noise levels. When the noise level was low and no chemicals loaded on multiple patterns, variational confidence intervals achieved 100\% coverage. As noise increased, coverage decreased, with added noise greater than or equal to $0.5 \times \sigma$ resulting in confidence intervals that did not achieve nominal coverage, i.e. did not include the true value 95\% of the time. As underlying patterns overlapped more, coverage again decreased, though notably less steeply than with increasing noise. At the extreme, with all patterns overlapping and added noise equal to $\sigma$, variational confidence intervals failed to include the true values almost 40\% of the time. This corresponds with poorer performance as the underlying structure became less separable. Variational confidence intervals consistently outperformed bootstrapped confidence intervals, detailed in Supplemental Section~\ref{supp_boot_results}.

\subsection{Application to a mixture of endocrine disrupting chemicals in pregnant women}
\label{results_app}
Primary data from pregnant women ages 18--35 were collected as part of the Columbia Center for Children's Environmental Health's Mothers and Newborns longitudinal birth cohort. This mother-child cohort was initiated to evaluate the effects of prenatal exposure to environmental contaminants on birth outcomes and child development \citep{perera2006effect}. Multiple potential EDCs were measured in spot urine samples collected during the third trimester in a subset of 343 mothers \citep{factor2014persistent}. Here, we evaluated a mixture of 17 chemicals from two classes: nine phthalate metabolites and eight phenols. Concentrations below the limit of detection (LOD) were assigned a value of LOD/$\sqrt{2}$ \citep{hornung1990estimation}. All chemical concentrations were specific gravity-adjusted to account for urinary dilution \citep{hauser2004temporal}. Exposures were scaled by their standard deviation but not mean-centered, to retain non-negativity.

Exposure levels of phthalates were generally positively correlated, though the majority of correlations (70\%) were weak (between 0.0 and 0.3). Four phthalate metabolites (MEHHP, MECPP, MEOHP, and MEHP) from the same parent compound had correlation coefficients $>$ 0.7. Phenols displayed lower within-class correlations, with the exception of 24-DCP and 25-DCP ($r$ = 0.94) and PPB and MPB ($r$ = 0.75). Notable between-class correlations include MEP's moderate correlation with two phenols ($r > 0.4$) and BPA's moderate correlations with four phthalates ($r > 0.2$).

As preliminary analyses, we fit PCA and NMF on the exposure concentrations. For PCA, we chose 11 principal components that explained 80\% of the variance in the data. For NMF, we chose a one factor model based on BIC. The identified factor loadings expressed common exposure to all chemicals, without distinguishing separate patterns.

\begin{figure}[!htbp]
\caption{Patterns of potentially endocrine disrupting chemical exposure in pregnant women identified by \bnmf in a mixture of 17 phenols and phthalates.}
\label{fig:patterns}
\centering
\includegraphics[scale = 0.3]{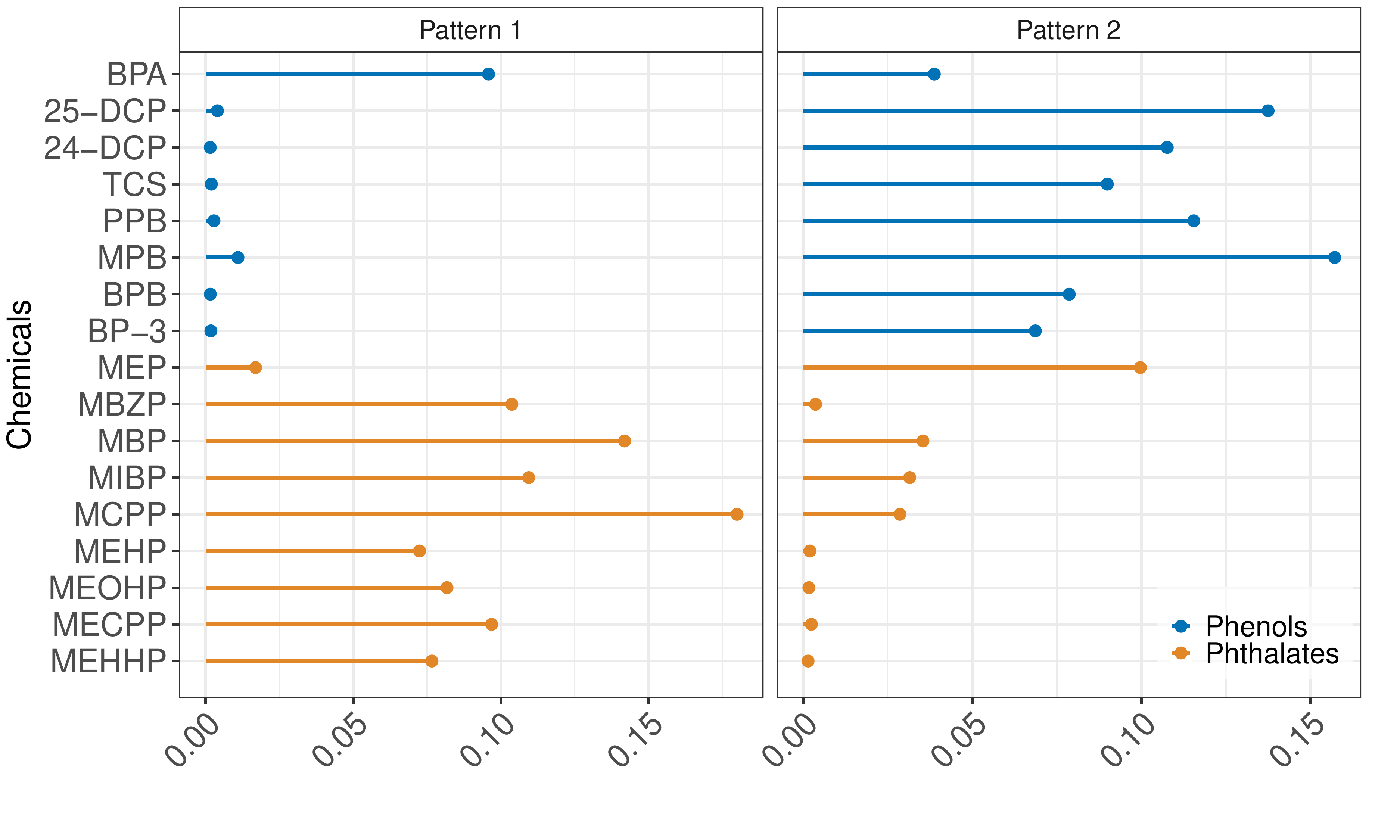}
\end{figure}

We then applied \bnmf to identify exposure patterns of phenols and phthalates in pregnant women without specifying the number of patterns. We ran 10 models using variational inference (described in Section \ref{methods_vi}), and chose the run with the largest objective value for inference. \bnmf identified two patterns of EDC exposure: (1) majority phthalates, and (2) majority phenols. Two notable cross-class loadings occurred; MEP loaded with phenols, and BPA loaded with phthalates. These groupings may be explained by potential shared predominant exposure routes through personal care products (phenols and MEP) and diet (phthalates and BPA). Pattern loadings are shown in Figure~\ref{fig:patterns}. 

\begin{figure}[!htbp]
\caption{Individual scores and 95\% variational confidence intervals on \bnmfc-identified patterns for a random subset of ten participants in the Mothers \& Newborns cohort. This shows variability across individual pattern scores and across patterns within individuals. Error bars span the upper and lower bounds of the 95\% variational confidence intervals, which display uncertainty in the estimation of individual scores.}
\label{fig:scores}
\centering
\includegraphics[scale = 0.5]{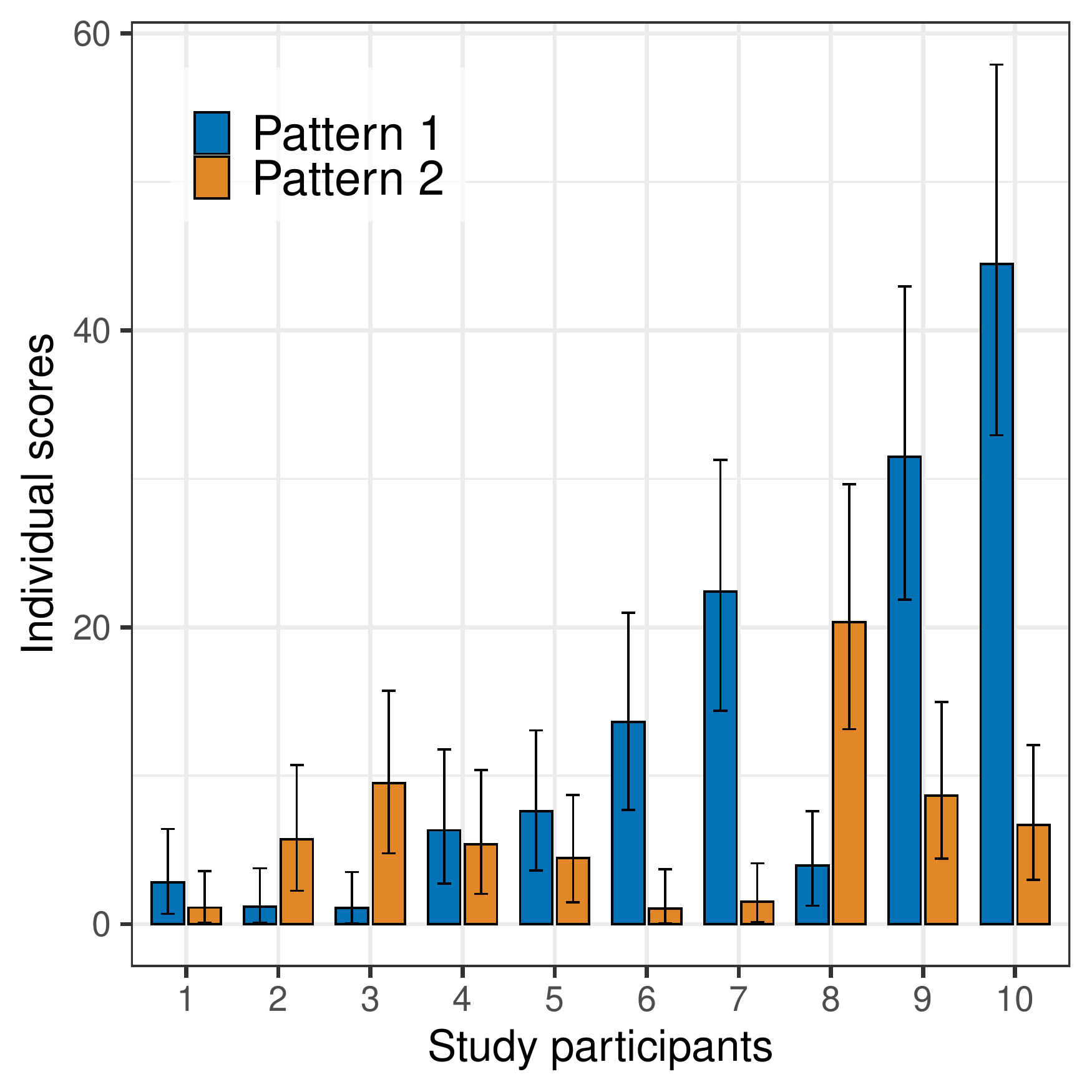}
\end{figure}

We also obtained variational confidence intervals around individual scores, which may be included in a health model. Each individual score had its own mean and variance. Score means across all individuals averaged 5.3 ng/ml (between person sd = 4.4) for pattern 1 and 4.7 ng/ml (sd = 3.4) for pattern 2, ranging from 1.1 to 44.5 ng/ml and 1.0 to 20.3 ng/ml, respectively. The average within-person variance of individual scores averaged 4.5 for pattern 1 and 3.9 for pattern 2, ranging from 0.9 to 38.6 and 0.9 to 17.0, respectively. We depict confidence intervals around a random subset of individual scores in Figure~\ref{fig:scores}.

\section{Discussion}

We have proposed \bnmf as a new approach to identify patterns of chemical exposures in environmental mixtures. Our simulation studies highlight two main advantages of \bnmf over existing frequentist approaches. First, by fitting a non-parametric prior on the number of patterns in a mixture, \bnmf helps to remove the researcher from the specification and selection of $K$. To our knowledge, all methods currently utilized in environmental health research require an \textit{a priori} specification or \textit{post-hoc} selection of pattern number. Second, the Bayesian framework allows us to estimate 95\% variational confidence intervals surrounding point estimates for individual scores. Coverage of true individual scores was $>$95\%, on average, when added noise was below 40\% in our simulations. Variational confidence intervals better capture uncertainty in inference and may aid researchers in the interpretation of results, especially if, for example, uncertainty is not the same across patterns or across individuals. Quantification of confidence may also be included in two-stage analyses, which could help in obtaining more comparable results across studies. Other Bayesian studies in environmental health have done something similar, modeling the spatio-temporal distribution of air pollution in the first stage, and estimating the corresponding health effects in the second stage while incorporating posterior uncertainty in the pollution predictions \citep{lee2017rigorous}. Health models that ignore this uncertainty may have overly narrow confidence intervals and unstable effect estimates, threatening inference \citep{kioumourtzoglou2014impact}.

In an application of \bnmf to real-life data, we found two patterns of EDCs in a cohort of pregnant women. While we do not know the true exposure sources or the data generating process in this example, the identified patterns can be interpreted as two exposure routes or sources. One pattern of exposure to seven phenols and one phthalate metabolite, MEP, aggregated EDCs whose primary exposure route is through personal care products. Phenols are used as plasticizers, antibacterial agents, and preservatives in a wide variety of personal care products \citep{philippat2015exposure}, and MEP is primarily found in fragrances \citep{koniecki2011phthalates}. The second pattern combined the remaining eight phthalates and BPA. The majority of these phthalates are metabolites of compounds found in food packaging \citep{schettler2006human}. BPA is also found in packaging, such as the lining of aluminium cans, and both phthalates and BPA are known to leach into food \citep{vandenberg2007human}. \bnmfc-identified variational confidence intervals further allowed for uncertainty characterization. For example, while variational confidence intervals exhibited substantial overlap across individuals, within-participant variances on patterns 1 and 2 were often quite disparate, indicating more confidence in one score than the other. This application shows \bnmf as a useful tool to identify exposure patterns in environmental mixtures. 

The primary and secondary simulations were implemented to describe a variety of potential contexts in which pattern recognition methods may be desired in environmental health research. We simulated settings ranging from distinct to overlapping patterns. We varied the number of patterns (1, 4, and 10), number of study subjects (200, 1,000, 10,000), and number of features (20, 40, and 100). We adjusted the inter-chemical correlations and the amount of added noise. \bnmf performed comparably with more commonly used methods on a number of metrics. However, there remain high-dimensional application areas, such as epigenetics or metabolomics, that are not well represented by our simulations. Computationally, the sample size and dimension of the exposure vector behave as limiting factors in these models, though the variational inference algorithm was written to enhance efficiency and may be further extended to a stochastic version to facilitate larger $N$ or $P$.

Bayesian methods may be sensitive to prior specifications. We chose Gamma priors on chemical loadings, individual scores, and the number of patterns because of their aptness in the setting of environmental concentrations. The Gamma distribution is non-negative and continuous (as are environmental exposures), with support $x \in(0, \infty)$, and finite mean and variance appropriate for real-life data. We specified $\operatorname{Gamma}(\alpha = \mathbf{1}, \beta = \mathbf{1})$ as a weakly informative distribution over our prior beliefs for the coefficient and dictionary matrices. This prior reflects a belief that is weakly held and easily molded by exposure to new information from empirical observations. We stipulated $\operatorname{Gamma}(\alpha = \frac{1}{K}, \beta = 1)$ to enforce sparsity on the number of patterns discovered.

Variational inference provides an accurate approximation of the full posterior distribution. Still, it is generally not recommended to infer properties of an underlying distribution beyond the expected value due to the difficulty in verifying the assumptions regarding the factorization of the posterior distribution. This begs the question, why not simulate \bnmfc's posterior distribution with a Markov chain Monte Carlo (MCMC)? Several considerations led us to prefer variational inference over MCMC for \bnmfc. First, we developed \bnmf primarily for use in environmental mixtures analyses, often performed by environmental epidemiologists. Variational inference is a more approachable method for many researchers. Pragmatically, by converting the inference problem into an optimization problem, variational inference outperforms MCMC in terms of analytic efficiency. An MCMC is more computationally intense and often assumes access to higher performance computing resources. Additionally, research on bounds for guaranteed MCMC mixing time is, in general, an open area \citep{levin2017markov}, which places the burden of assessing convergence on the end user. This is not a trivial task, especially for users without formal training, and variational inference avoids it by converging to a local minimum. Multiple runs of variational inference are easily ranked in terms of performance; the model that obtains the highest objective value is the best approximation of the true posterior distribution. In this way, variational inference is more widely usable. 

A second consideration was the complexity of the model at hand. Because of the nonparametric prior on the number of patterns in $\mathbf{a}$, \bnmf is not conjugate, thus requiring a more complex MCMC than a Gibbs sampler. This, in itself, did not make an MCMC impractical, but the nonparametric prior created an additional difficulty wherein samples within the same chain could result in different numbers of patterns, complicating the Monte Carlo step.

An MCMC would further suffer from label switching, which makes the posterior distribution non-identifiable. The posterior distribution is invariant to permutations of the labels of patterns $1 ... K$. Label switching occurs when the pattern labels permute \citep{celeux1998bayesian}, resulting in an uninterpretable posterior. While there are certain proposed approaches to address this issue, such as parameter ordering constraints or relabeling strategies \citep{rodriguez2014label}, there are no consensus solutions \citep{gelman06}. Variational inference sidesteps label switching though iterative optimization; the patterns inferred in the first iteration initialize the next iteration, resulting in a sequence of improving approximations.

\begin{figure}[!htbp]
\caption{Ninety-five percent variational confidence interval coverage. Each square represents 100 simulated datasets colored according to median coverage (proportion of true values within estimated 95\% variational confidence intervals). On the x axis, number of distinct chemicals per pattern ranges from 10 (distinct) to 0 (overlapping). On the y axis, added noise level relative to the true standard deviation increases from 0 (no added noise) to 1 (as much added noise as true standard deviation).}
\label{heatmap}
\centering
\includegraphics[scale = 0.65]{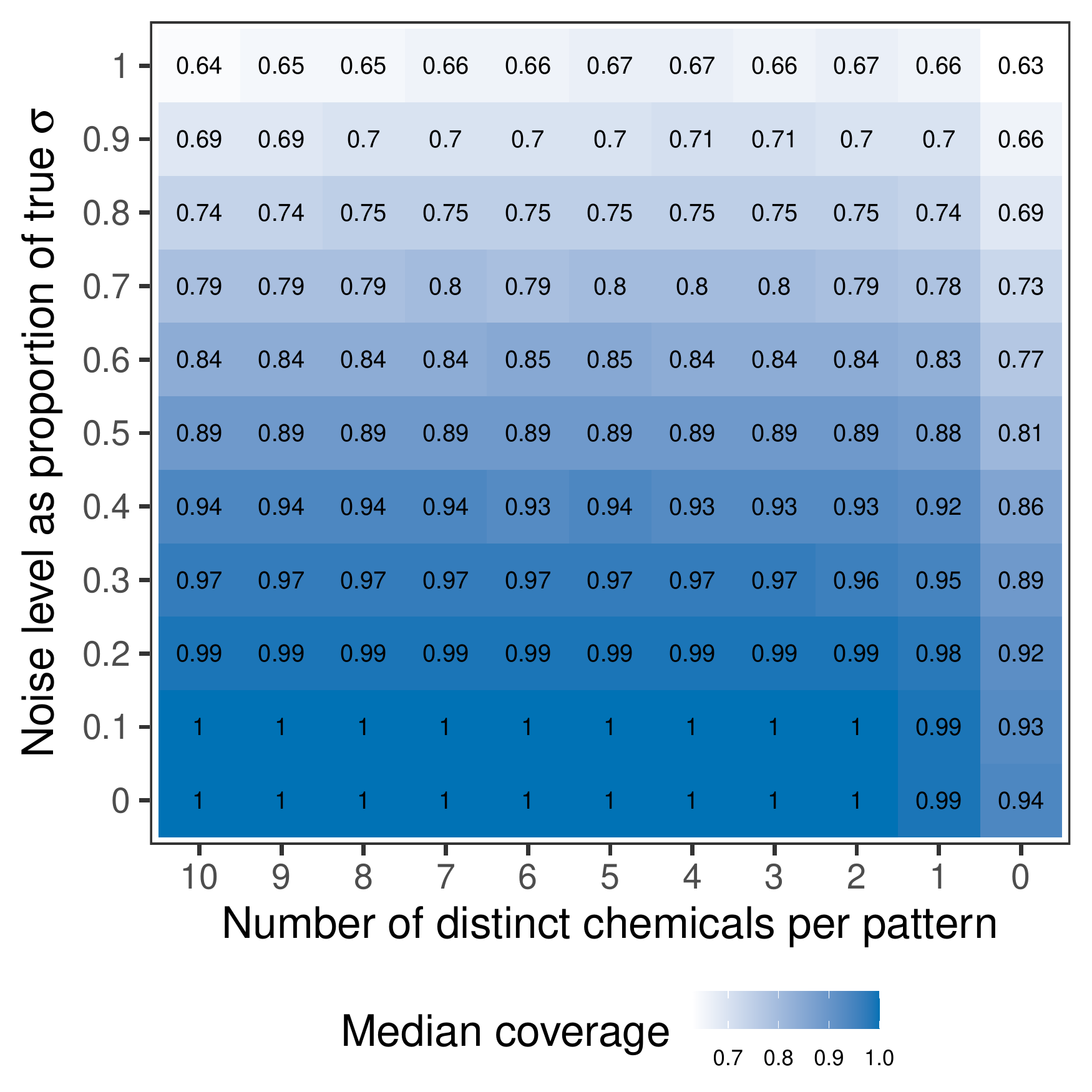}
\end{figure}

We show in Figure~\ref{heatmap} that as long as separability holds, \bnmfc's variational confidence intervals achieves nominal coverage or better. In fact, even when patterns were assumed to be known (i.e. with zero error in dictionary matrix estimation), both the error in coefficient matrix estimation and variational confidence interval coverage decreased as noise was added (results not shown), indicating that declines in performance were due to a loss of separability and not to a deficiency of variational inference. In fact, variational confidence intervals consistently performed better than bootstrapped confidence intervals, which never achieved nominal coverage.

In Section~\ref{methods_bayesian} we described the probabilistic interpretation on which \bnmf was built. It began with modeling the data as independent random variables and drawing latent variables independently from Gamma priors \citep{cemgil2008bayesian, paisley2014bayesian}. Variational inference, likewise, models latent variables as mutually independent and governed by a distinct factor in the variational density \citep{blei2017variational}. In this way, assumptions of variational inference cohere with assumptions of the data generating process of all NMF methods, including \bnmf.

\bnmf performed comparably to other pattern identification methods in simulations across several metrics of error in magnitude of values and direction of patterns. \bnmf and frequentist NMF have the advantage over factor analysis and PCA of returning solution matrices on immediately interpretable scales. For example, in our EDC application, chemicals were measured in nanograms of a chemical per milliliter of urine. After non-negative decomposition, individual scores were represented as nanograms of a pattern per milliliter of urine, and chemical loadings were represented as nanograms of a chemical per nanogram of a pattern. PCA and factor analysis have no such intuitive framing.

\bnmf had additional benefits over frequentist NMF. Although they performed similarly in simulations, when we applied them to real-life data and used BIC to choose the optimal model, both NMF with an $\ell_2$ penalty and with a Poisson likelihood returned a single factor. Since the goal of the analysis was to identify patterns that could be interpreted as chemical sources and not to create an index of chemical exposures, a single factor failed to address the research question. Additionally, bootstrapped confidence intervals around individual scores from NMF (see Supplemental Figure~\ref{fig:boot_coverage}) performed substantially worse than variational confidence intervals, even when simulated noise was low, indicating that the posterior distribution of \bnmf better encapsulate the truth than the maximum likelihood analog.

With this work we aimed to improve reproducibility and robustness of pattern identification in environmental health. We incorporated a non-parametric sparse prior in a Bayesian model to estimate the number of patterns in high-dimensional environmental mixtures. This approach may aid researchers in model selection. Additionally, the Bayesian formulation allows for uncertainty quantification. \bnmf may be used as a feature selection step in a Bayesian two-stage analysis incorporating individual pattern scores and distributions as exposures in subsequent health models. Our model achieved nominal coverage even with moderate noise. In future work, we will build the hierarchical Bayesian health model and consider useful extensions, such as 1) a fully supervised model where the health outcome informs pattern identification, and 2) accounting for clustered information, e.g. repeated measurements or geographical clustering.

\begingroup
\renewcommand{\arraystretch}{1.5}
\begin{table}[!htbp] \centering 
  \caption{Relative error comparing predicted values and solution scores and loadings with simulated truth. Values presented are mean (standard deviation). Entries marked as ``---'' indicate that no models on that group of simulations correctly identified the number of patterns, and error on scores and loadings could not be computed. $^*$ FA = factor analysis; NMF-$\ell_2$ = NMF with $\ell_2$ penalty; NMF-P = NMF with Poisson likelihood.} 
  \label{tab_l2} 
  \addtolength{\tabcolsep}{-2pt}
\small
\begin{tabular}{lccc|ccc}
\multicolumn{7}{c}{Relative Prediction Error} \\
\hline 
\hline  
& \multicolumn{3}{c}{Distinct Patterns} & \multicolumn{3}{c}{Overlapping Patterns} \\
\hline
\hline  
Model$^*$ & Overall & Scores & Loadings & Overall & Scores & Loadings \\ 
\hline
\hline  
& \multicolumn{6}{c}{Simulations + 20\% Noise} \\
 \hline 
\bnmf & 0.07 ($<$0.01) & 0.77 (0.03) & 3.18 (1.07) & 0.07 ($<$0.01) & 0.78 (0.01) & 3.48 (0.38) \\ 
FA & 0.43 (0.03) & 0.75 (0.01) & 0.02 ($<$0.01) & 0.29 (0.04) & 0.78 (0.01) & 0.52 (0.03) \\ 
NMF-$\ell_2$ & 0.05 ($<$0.01) & 1.00 ($<$0.01) & 1586.63 (40.82) & 0.04 ($<$0.01) & 1.00 ($<$0.01) & 1623.25 (37.28) \\
NMF-P & 0.05 ($<$0.01) & 10.98 (3.69) & 0.91 (0.02) & 0.05 ($<$0.01) & 8.81 (2.99) & 0.89 (0.03) \\ 
PCA & 0.09 (0.10) & 1.89 (0.07) & 0.72 (0.03) & 0.18 (0.02) & --- & --- \\ 
 \hline 
& \multicolumn{6}{c}{Simulations + 50\% Noise} \\
 \hline 
\bnmf & 0.14 ($<$0.01) & 0.77 (0.03) & 3.37 (1.28) & 0.12 (0.02) & 0.79 (0.01) & 3.61 (0.28) \\ 
FA & 0.47 (0.03) & 0.75 ($<$0.01) & 0.09 ($<$0.01) & 0.33 (0.03) & 0.78 (0.01) & 0.46 (0.03) \\ 
NMF-$\ell_2$ & 0.13 ($<$0.01) & 1.00 ($<$0.01) & 1654.97 (49.45) & 0.11 ($<$0.01) & 1.00 ($<$0.01) & 1684.69 (41.46) \\ NMF-P & 0.13 ($<$0.01) & 10.89 (3.50) & 0.91 (0.02) & 0.11 ($<$0.01) & 8.62 (2.86) & 0.89 (0.03) \\ 
PCA & 0.13 ($<$0.01) & 1.86 (0.06) & 0.72 (0.03) & 0.11 (0.01) & 1.62 (0.06) & 0.79 (0.03) \\ 
 \hline 
& \multicolumn{6}{c}{Simulations + 100\% Noise} \\
 \hline 
\bnmf & 0.27 (0.01) & 0.83 (0.06) & 5.72 (2.90) & 0.23 (0.01) & 0.81 (0.02) & 4.22 (0.70) \\ 
FA & 0.56 (0.03) & 0.77 ($<$0.01) & 0.24 ($<$0.01) & 0.43 (0.03) & 0.80 (0.01) & 0.39 (0.02) \\ 
NMF-$\ell_2$ & 0.27 (0.01) & 1.00 ($<$0.01) & 1850.65 (61.52) & 0.22 (0.01) & 1.00 ($<$0.01) & 1873.28 (60.33) \\ 
NMF-P & 0.27 (0.01) & 10.06 (2.50) & 0.90 (0.02) & 0.23 (0.01) & 8.08 (2.59) & 0.88 (0.03) \\ 
PCA & 0.48 (0.01) & --- & --- & 0.45 (0.01) & --- & --- \\ 
\hline
\hline  
\end{tabular}
\end{table}
\endgroup

\begingroup
\renewcommand{\arraystretch}{1.5}
\begin{table}[!htbp] \centering 
  \caption{Cosine distance comparing predicted values and solution scores and loadings with simulated truth. Values represent mean (standard deviation). Entries marked as ``---'' indicate that no models on that group of simulations correctly identified the number of patterns, and distance on scores and loadings could not be computed. $^*$ FA = factor analysis; NMF-$\ell_2$ = NMF with $\ell_2$ penalty; NMF-P = NMF with Poisson likelihood.} 
  \label{tab_cos}
  \addtolength{\tabcolsep}{-2pt}
\small
\begin{tabular}{lccc|ccc}
\multicolumn{7}{c}{Cosine Distance} \\
\hline 
\hline  
& \multicolumn{3}{c}{Distinct Patterns} & \multicolumn{3}{c}{Overlapping Patterns} \\
\hline
\hline  
Model$^*$ & Overall & Scores & Loadings & Overall & Scores & Loadings \\ 
\hline
\hline  
& \multicolumn{6}{c}{Simulations + 20\% Noise} \\
 \hline 
\bnmf & $<$0.01 ($<$0.01) & $<$0.01 ($<$0.01) & $<$0.01 ($<$0.01) & $<$0.01 ($<$0.01) & 0.02 (0.01) & 0.02 (0.01)  \\
FA & 0.08 (0.01) & 0.21 (0.01) & $<$0.01 ($<$0.01) & 0.03 (0.01) & 0.28 (0.01) & 0.08 (0.01) \\
NMF-$\ell_2$ & $<$0.01 ($<$0.01) & $<$0.01 ($<$0.01) & $<$0.01 ($<$0.01) & $<$0.01 ($<$0.01) & $<$0.01 ($<$0.01) & $<$0.01 ($<$0.01 \\
NMF-P & $<$0.01 ($<$0.01) & $<$0.01 ($<$0.01) & $<$0.01 ($<$0.01) & $<$0.01 ($<$0.01) & $<$0.01 ($<$0.01) & $<$0.01 ($<$0.01) \\
PCA & 0.01 (0.02) & 0.28 (0.06) & 0.09 (0.07) & 0.02 ($<$0.01) & --- & ---  \\
 \hline 
& \multicolumn{6}{c}{Simulations + 50\% Noise} \\
 \hline 
\bnmf & 0.01 ($<$0.01) & 0.01 ($<$0.01) & $<$0.01 ($<$0.01) & 0.01 ($<$0.01) & 0.02 (0.01) & 0.02 (0.01) \\
FA & 0.10 (0.02) & 0.22 (0.01) & $<$0.01 ($<$0.01) & 0.05 (0.01) & 0.29 (0.02) & 0.08 (0.01) \\
NMF-$\ell_2$ & 0.01 ($<$0.01) & 0.01 ($<$0.01) & $<$0.01 ($<$0.01) & 0.01 ($<$0.01) & 0.01 ($<$0.01) & $<$0.01 ($<$0.01) \\
NMF-P & 0.01 ($<$0.01) & 0.01 ($<$0.01) & $<$0.01 ($<$0.01) & 0.01 ($<$0.01) & 0.01 ($<$0.01) & $<$0.01 ($<$0.01) \\
PCA & 0.01 ($<$0.01) & 0.28 (0.06) & 0.08 (0.06)  & 0.01 ($<$0.01) & 0.48 (0.05) & 0.35 (0.05) \\
 \hline 
& \multicolumn{6}{c}{Simulations + 100\% Noise} \\
 \hline 
\bnmf & 0.03 ($<$0.01) & 0.03 ($<$0.01) & $<$0.01 ($<$0.01) & 0.02 ($<$0.01) & 0.05 (0.01) & 0.03 (0.01) \\
FA & 0.17 (0.02) & 0.24 (0.01) & $<$0.01 ($<$0.01) & 0.10 (0.01) & 0.31 (0.02) & 0.08 (0.01) \\
NMF-$\ell_2$ & 0.03 ($<$0.01) & 0.03 ($<$0.01) & $<$0.01 ($<$0.01) & 0.02 ($<$0.01) & 0.04 ($<$0.01) & 0.01 ($<$0.01) \\
NMF-P & 0.03 ($<$0.01) & 0.03 ($<$0.01) & 0.01 ($<$0.01) & 0.02 ($<$0.01) & 0.04 ($<$0.01) & 0.01 ($<$0.01) \\
PCA & 0.09 ($<$0.01) & --- & --- & 0.08 ($<$0.01) & --- & --- \\
\hline 
\hline
\end{tabular}
\end{table}
\endgroup

\begingroup
\renewcommand{\arraystretch}{1.5}
\begin{table}[!htbp] \centering 
  \caption{Symmetric subspace distance comparing predicted values and solution scores and loadings with simulated truth. Values represent mean (standard deviation). $^*$ FA = factor analysis; NMF-$\ell_2$ = NMF with $\ell_2$ penalty; NMF-P = NMF with Poisson likelihood.} 
  \label{tab_ssd}
  \addtolength{\tabcolsep}{-2pt}
\begin{tabular}{lccc|ccc}
\multicolumn{7}{c}{Symmetric Subspace Distance} \\
\hline 
\hline  
& \multicolumn{3}{c}{Distinct Patterns} & \multicolumn{3}{c}{Overlapping Patterns} \\
\hline
\hline  
Model$^*$ & Overall & Scores & Loadings & Overall & Scores & Loadings \\ 
\hline
\hline  
& \multicolumn{6}{c}{Simulations + 20\% Noise} \\
\hline
\bnmf & 0.03 (0.02) & 0.06 ($<$0.01) & 0.01 ($<$0.01) & 0.05 (0.02) & 0.08 (0.01) & 0.06 (0.01) \\ 
FA & 0.16 (0.01) & 0.43 (0.01) & 0.01 ($<$0.01) & 0.17 (0.01) & 0.43 (0.01) & 0.07 (0.01) \\ 
NMF-$\ell_2$ & 0.02 (0.02) & 0.06 ($<$0.01) & 0.01 ($<$0.01) & 0.05 (0.02) & 0.08 ($<$0.01) & 0.01 ($<$0.01) \\ 
NMF-P & 0.02 (0.02) & 0.06 ($<$0.01) & 0.01 ($<$0.01) & 0.05 (0.02) & 0.08 ($<$0.01) & 0.02 ($<$0.01) \\ 
PCA & 0.16 (0.01) & 0.45 (0.07) & 0.09 (0.18) & 0.17 (0.01) & 0.65 ($<$0.01) & 0.50 ($<$0.01) \\ 
\hline
& \multicolumn{6}{c}{Simulations + 50\% Noise} \\
\hline
\bnmf & 0.05 (0.01) & 0.14 ($<$0.01) & 0.02 ($<$0.01)& 0.08 (0.02) & 0.18 (0.05) & 0.08 (0.06) \\
FA & 0.17 (0.01) & 0.44 ($<$0.01) & 0.02 ($<$0.01) & 0.18 (0.01) & 0.46 (0.01) & 0.08 (0.01) \\ 
NMF-$\ell_2$ & 0.05 (0.01) & 0.13 ($<$0.01) & 0.02 ($<$0.01)& 0.08 (0.02) & 0.18 (0.01) & 0.03 ($<$0.01) \\ 
NMF-P & 0.05 (0.01) & 0.13 ($<$0.01) & 0.03 ($<$0.01) & 0.08 (0.02) & 0.18 (0.01) & 0.05 (0.01) \\
PCA & 0.17 ($<$0.01) & 0.44 ($<$0.01) & 0.03 ($<$0.01) & 0.18 (0.01) & 0.46 (0.03) & 0.05 (0.08) \\ 
\hline
& \multicolumn{6}{c}{Simulations + 100\% Noise} \\
\hline
\bnmf & 0.10 (0.04) & 0.29 (0.09) & 0.10 (0.15) & 0.11 (0.01) & 0.31 (0.02) & 0.11 (0.04) \\ 
FA    & 0.18 (0.01) & 0.49 (0.01) & 0.05 ($<$0.01) & 0.20 (0.01) & 0.53 (0.01) & 0.10 (0.01) \\ 
NMF-$\ell_2$ & 0.09 (0.03) & 0.26 (0.06) & 0.07 (0.10) & 0.12 (0.01) & 0.31 (0.01) & 0.07 (0.01) \\ 
NMF-P  & 0.09 (0.02) & 0.26 (0.04) & 0.07 (0.07) & 0.11 (0.01) & 0.32 (0.01) & 0.11 (0.01) \\ 
PCA   & 0.60 (0.02) & 0.91 (0.01) & 0.88 (0.01) & 0.62 (0.02) & 0.92 ($<$0.01) & 0.88 (0.01) \\ 
\hline
\hline
\end{tabular}
\end{table}
\endgroup

\clearpage
\section{Supplementary Material}
The supplemental materials include detailed information about (1) metrics and measures of comparison, (2) variational inference, (3) bootstrapped confidence interval creation and resulting comparison with variational confidence intervals, and (4) secondary analyses varying $N$, $P$, and $K$.

\section*{Funding}
This work was supported by the National Institutes of Environmental Health (NIEHS) individual fellowship grant F31 ES030263, as well as PRIME R01 ES028805 and P30 ES009089.

\section*{Acknowledgement}
\textit{Conflict of Interest:} None declared.

\bibliographystyle{biorefs.bst}
\bibliography{main.bib}

\clearpage
\setcounter{figure}{0}
\setcounter{table}{0}
\setcounter{section}{4}

\renewcommand{\thefigure}{S\arabic{figure}}
\renewcommand{\thetable}{S\arabic{table}}

\section{Supplementary Material}
\label{sec6}

\subsection{Detailed metrics and measures of comparison}
\label{supp_metrics}

As a first qualitative metric, we compared the number of patterns identified. As quantitative metrics, we used relative predictive error, cosine distance, and symmetric subspace distance. We used the relative predictive error in terms of the Frobenius norm ($\ell_2$ matrix norm) to compare solution matrices to simulated datasets before noise was added, which represent the true underlying mixture. Relative error is defined as the ratio of the error to the truth: $\left\lVert Truth - Predicted\right\rVert_F / \left\lVert  Truth\right\rVert_F$. 

For solutions that correctly identified the true number of patterns, we used relative error and cosine distance to compare estimated coefficient matrices or scores to true simulated scores and estimated dictionary matrices, loadings, or factors to true simulated loadings. Cosine distance measures the distance between two vectors in terms of orientation, not magnitude. It is defined as 1 -- cosine similarity, where cosine similarity is the inner product of two vectors, $\mathbf{a}$ and $\mathbf{b}$, which are both normalized so that $\|\mathbf{a}\|_2 = 1$ and $\|\mathbf{b}\|_2 = 1$ \citep{tan2016introduction}. 
\begin{equation}
\text{cosine distance}(\mathbf{a}, \mathbf{b}) = 1 - \cos (\theta)=1 - \frac{\mathbf{a} \cdot \mathbf{b}}{\|\mathbf{a}\|_2\|\mathbf{b}\|_2},
\end{equation}

\noindent where $\theta$ is the angle between $\mathbf{a}$ and $\mathbf{b}$. With cosine distance, intuitively, smaller values indicate more similarity, with a value of 0 demonstrating that the two vectors are oriented in the same direction ($1 - \cos(0)$) and 1 indicating orthogonality.

As solution factors or components are not guaranteed to match the order of the true underlying patterns, for each solution, we found the permutation matrix that made the solution matrix as close to the true matrix as possible, finding a globally optimal correspondence between the true matrix and the solution. We then reordered the solution matrix by the permutation matrix. Because PCA and factor analysis solutions are unique up to sign, we allowed their permutation matrices to include negative values.

Because all solutions did not successfully identify the correct number of patterns, we utilized symmetric subspace distance to compare  matrices of different dimensions (e.g., true coefficient matrix $1,000 \times 4$ vs. estimated coefficient matrix $1,000 \times 5$). Symmetric subspace distance defines a distance measure between two linear subspaces, where the number of individual vectors is constant but the number of elements they contain may differ \citep{wang2006subspace}. The symmetric distance between $m$-dimensional subspace $U$ and $n$-dimensional subspace $V$ is defined as 

\begin{equation}
d(U, V) =\sqrt{\max (m, n)-\sum_{i=1}^{m} \sum_{j=1}^{n}\left(u_{i}^{\mathrm{T}} v_{j}\right)^{2}}
\end{equation}

\noindent If two subspaces largely overlap, they will have a small distance; if they are almost orthogonal, they will have a large distance. To use this method to compare simulations with solutions, we took orthonormal bases of all included matrices and calculated the ratio between symmetric subspace distance and $\sqrt{max(m,n)}$ \citep{wang2006subspace}. This method may favor results from PCA, which provides orthonormal bases directly. It also removes non-negativity as a characteristic of either simulations or solutions, thus it may artificially improve performance of methods that allow negative values.

\subsection{Variational inference}
\label{suppVI}  
Here, we defined the variational approximating distributions $q(W)$, $q(\mathbf{a})$, and $q(H)$ as follows:
\begin{equation}
q(W) =\operatorname{Gamma}\left(W_{1}, W_{2}\right); \hspace{1ex} q(\mathbf{a})=\operatorname{Gamma}\left(\mathbf{a}_{1}, \mathbf{a}_{2}\right); \hspace{1ex}
q(H) =\operatorname{Gamma}\left(H_{1}, H_{2}\right)
\end{equation}

\noindent Given a dataset $X$, variational inference chooses variational parameters that minimize the Kullback-Liebler (KL) divergence between the variational family and the exact posterior \citep{jordan1999introduction}, where $\lambda$ includes all variational parameters, 
\begin{equation}
    \boldsymbol{\lambda}^{*}=\arg \min \mathrm{KL}(q(W, \mathbf{a}, H ; \boldsymbol{\lambda}) \| p(W, \mathbf{a}, H \mid X)).
\end{equation}

\noindent KL divergence is defined as,
\begin{equation}
\mathrm{KL}(q(W, \mathbf{a}, H ; \boldsymbol{\lambda}) \| p(W, \mathbf{a}, H | X))=\mathbb{E}_{q}\left[\log \frac{q(W, \mathbf{a}, H ; \boldsymbol{\lambda})}{p(W, \mathbf{a}, H | X)}\right]
\end{equation} 

\noindent where $p(\cdot)$ is the posterior distribution over the parameters and $q(\cdot)$ is the distribution over the variational parameters. We cannot actually calculate the KL divergence because it includes the posterior itself, which is intractable. Instead of optimizing the KL divergence directly, variational inference maximizes the negative KL divergence plus $\log p(x)$, which is a constant with respect to $q(\cdot)$. Maximizing this variational objective function is equivalent to minimizing the KL divergence \citep{blei2017variational}. We defined the variational objective function as follows:
\begin{equation}
\mathcal{L}=\mathbb{E}_{q}\left[\log \frac{p(W)}{q(W)}\right]+\mathbb{E}_{q}\left[\log\frac{p(\mathbf{a})}{q(\mathbf{a})}\right]+\mathbb{E}_{q}\left[\log\frac{p\left(X \mid W, \mathbf{a}, H\right) p(H)}{q(H)}\right]
\end{equation}

\noindent All three terms are expectations with respect to the variational distribution. We used the mean-field variational family where each latent variable is independent and governed by its own variational parameter \citep{jordan1999introduction, paisley2014bayesian}. The algorithm for learning the parameters of these $q$ distributions was provided in the original implementation of this method by \citet{holtzman2018machine}.

We have also included a deterministic annealing step to provide better approximations of the posterior distribution. This approach was first proposed to smooth the objective function and avoid local minima. It includes a decreasing temperature parameter that deforms the objective function over the course of the optimization \citep{rose1990deterministic}. With the inclusion of the temperature parameter, $T$, the objective function becomes:
\begin{equation}
\mathcal{L} = \mathbb{E}_{q}
\left[\log\left(p(X \mid W, \mathbf{a}, H)p(W)p(\mathbf{a})p(H)\right)\right] - T \times
\mathbb{E}_{q}\left[\log\left(q(W)q(\mathbf{a})q(H)\right)\right]
\end{equation}

\noindent When $T > 1$, the entropy is encouraged to be larger because low entropy distributions are penalized more \citep{mandt2016variational}. This, in practice, inflates the variance around variational parameters, better accounting for uncertainty in estimation. We have included an annealing schedule that shrinks $T$ to one in an iteration-dependent manner.

Variational inference is generally a non-convex optimization problem, thus it converges to a local, not global, maximum \citep{wainwright2008graphical, blei2017variational}. We included the deterministic annealing step described above to address the non-convexity. We also ran \bnmf 10 times for each simulated dataset and selected the version with the largest objective function, a standard practice that corresponds with selecting the variational distribution closest to the true posterior \citep{tran2021practical}. After convergence, we took $\mathbb{E}_{q}\left[W_{i k}\mathbf{a}_{k}\right]$ and $\mathbb{E}_{q}\left[H_{k j}\right]$ as individual scores and chemical loadings, respectively.

\subsection{Bootstrapped confidence intervals}
\label{supp_boot}

The variance of a parameter obtained through variational inference is narrower than the variance of the true posterior distribution as a consequence of the independence assumptions of the variational objective function and the clear trade-off between variational distributions that fit the data well and variational distributions that have low entropy \citep{blei2017variational}. 

We recognize that variational confidence intervals and bootstrapped confidence intervals are not directly comparable. The bootstrap forms a confidence interval over the range of expected values across bootstraps. This captures uncertainty in the data. The variational confidence interval captures the uncertainty in the variational approximation to the posterior distribution; this incorporates both uncertainty in the data and in the model. As the field currently stands, however, the best method of capturing uncertainty in pattern recognition in environmental mixtures is bootstrapping a frequentist model. Thus bootstrapped confidence intervals for \bnmf provide a measure against which to evaluate variational confidence intervals. Because a Bayesian estimate is not necessarily equivalent to the maximum likelihood estimate, for further comparison, we bootstrapped confidence intervals for the Poisson NMF, which, as discussed above, is the maximum likelihood corollary of \bnmfc.

We designed the bootstrap in the following manner: for a single simulation, we conducted 150 bootstraps using case resampling with replacement, so that each individual would appear in approximately 100 samples. The dimensions of each sample equaled the size of the original dataset. We then ran \bnmf on each bootstrapped sample and took $\mathbb{E}\left[W\operatorname{diag}(\mathbf{a})\right]$ and $\mathbb{E}\left[H\right]$ as point estimates. We employed the same scaling steps as above, $\ell_1$-normalization of the dictionary matrix and corresponding scaling of the coefficient matrix, on each bootstrap sample. We combined samples according to their row and column indices. We defined the 95\% bootstrapped confidence intervals as the 2.5\textsuperscript{th} and 97.5\textsuperscript{th} quantiles of the of the bootstrapped distribution of the scaled scores. We bootstrapped a subset of the simulated datasets detailed in Section~\ref{methods_sim}. We performed the same steps to create bootstrapped confidence intervals for Poisson NMF, included in the supplemental materials.

\subsubsection{Results for bootstrapped confidence intervals.}
\label{supp_boot_results}

While variational confidence intervals only performed poorly in simulations with high noise, bootstrapped confidence intervals consistently performed much worse. Median coverage for bootstrapped \bnmf confidence intervals ranged from 0.06 at its worst (over distinct patterns and low noise) to 0.19 at its best (over overlapping patterns and low noise). Median coverage for bootstrapped confidence intervals is included in Supplemental Figure~\ref{fig:boot_coverage}. \bnmfc's variational confidence intervals were wider than bootstrapped confidence intervals 97\% of the time, indicating that \bnmf appropriately accounted for uncertainty in the coefficient matrix. The narrower bootstrapped confidence intervals imply that less uncertainty was due to randomness in the simulated data than in the model itself.

\begin{figure}[h]
\caption{Example distributions for variational and bootstrapped 95\% confidence intervals of \bnmf and bootstrapped 95\% confidence intervals of NMF with the Poisson likelihood. Histograms show full distributions for these three distributions over a single entry in each solution matrix, i.e., an individual's estimated score on one pattern ($\mathbb{E}[W_{i k}\mathbf{a_k}]$), a chemical's estimated loading on one pattern ($\mathbb{E}[H_{k j}]$), and an individual's predicted value for one chemical ($\mathbb{E}[W_{i k}\mathbf{a}_k]*\mathbb{E}[H_{k j}]$). Dashed black lines represent the true value. Dashed colored lines represent the variational mean (blue), bootstrapped \bnmf median (red), and bootstrapped NMF median (green). Dotted colored lines represent the 95\% confidence intervals.}
\label{fig:ci_dist}
\centering
\includegraphics[scale = 0.55]{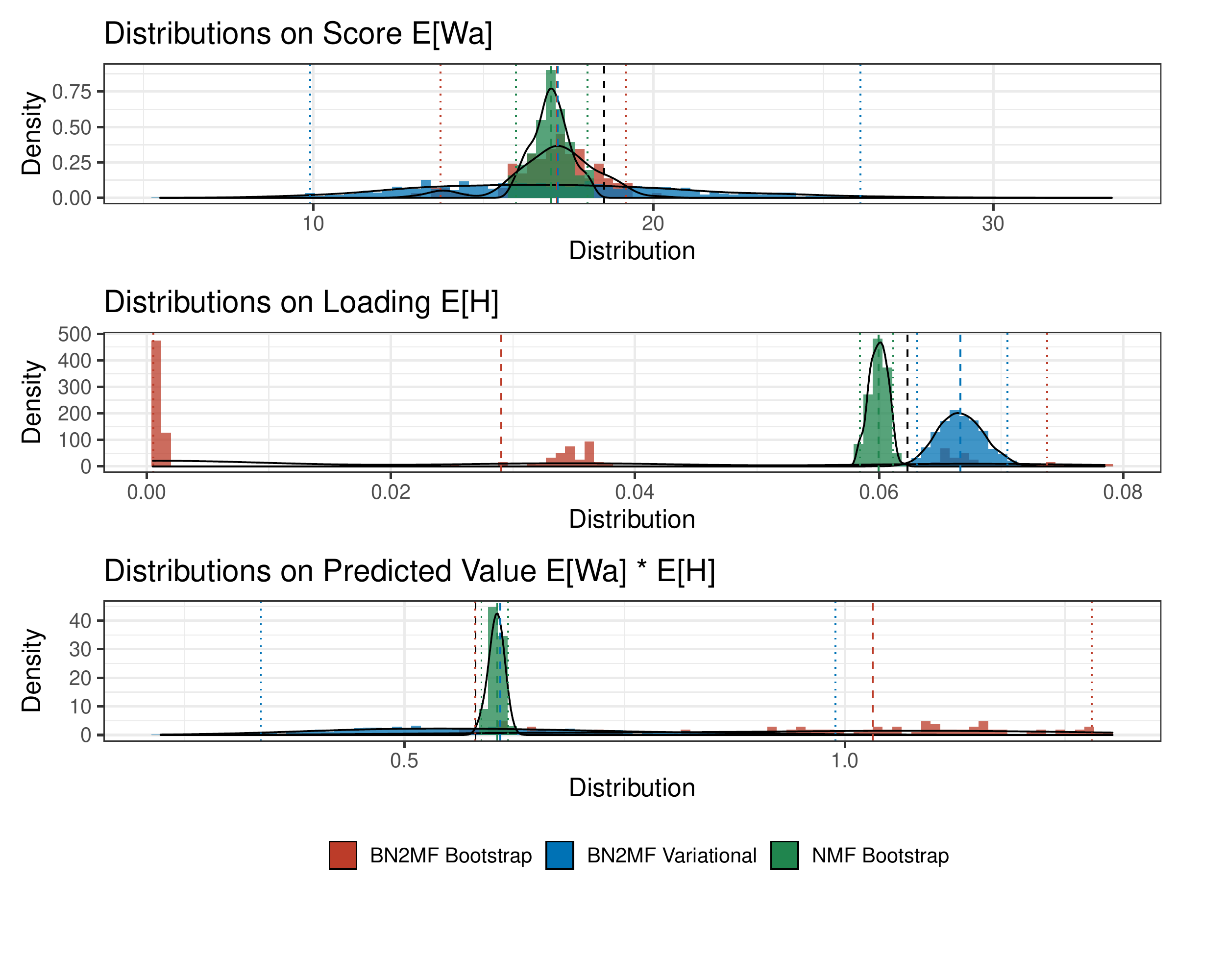}
\end{figure}

Variational confidence intervals derive from the distribution of $\hat{W}\operatorname{diag}(\hat{\mathbf{a}}) \sim \operatorname{Gamma}(\hat{\alpha}_W, \hat{\beta}_W) \times \\ \operatorname{Gamma}(\hat{\alpha}_\mathbf{a}, \hat{\beta}_\mathbf{a})$, which notably includes both tails of the distribution, whereas bootstrapped confidence intervals stem from the distribution of $\mathbb{E}\left[\hat{W}\operatorname{diag}(\hat{\mathbf{a}})\right]$ across bootstraps. Bootstraps generated similar expected values, which resulted in necessarily narrower confidence intervals. Bootstrapping, however, did provide wider confidence intervals for chemical loadings 85\% of the time. Supplemental figure~\ref{fig:ci_dist} gives representative examples of differences between variational and bootstrapped confidence intervals. We additionally included comparisons with bootstrapped confidence intervals for Poisson NMF in Supplemental Figures~\ref{fig:ci_dist} and \ref{fig:boot_coverage}. 

\begin{figure}
\caption{Ninety-five percent confidence interval coverage for variational confidence intervals, bootstrapped \bnmf confidence intervals, and bootstrapped Poisson NMF confidence intervals. Each square represents simulation that estimated a 4 pattern solution, colored according to median coverage (proportion of true values within estimated 95\% variational confidence intervals). On the x axis, number of distinct chemicals per pattern is 10 (distinct) or 0 (overlapping). On the y axis, added noise level relative to the true standard deviation increases from 0.2 (low added noise) to 1 (as much added noise as true standard deviation).}
\label{fig:boot_coverage}
\centering
\includegraphics[scale = 0.7]{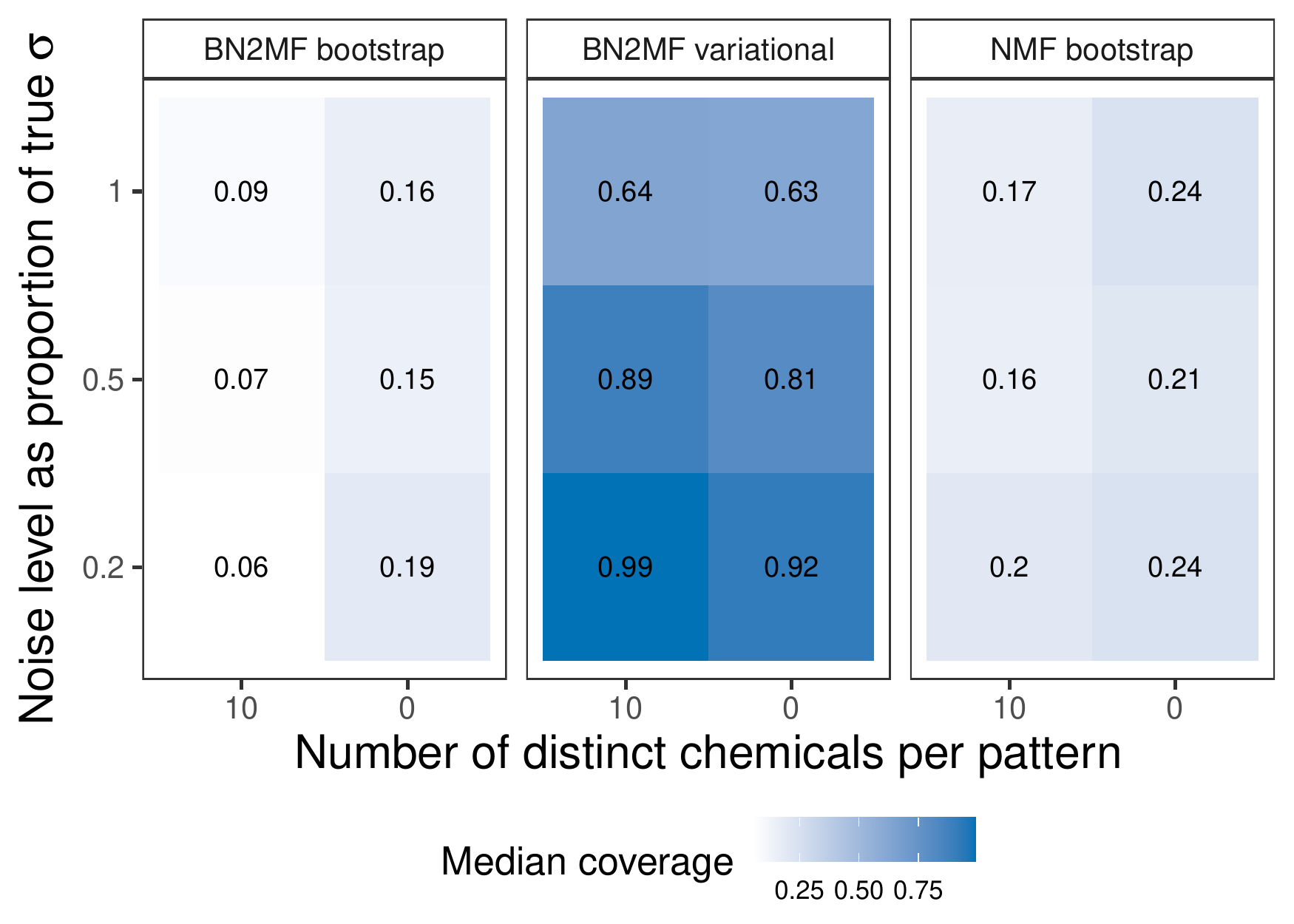}
\end{figure}

\clearpage
\subsection{Secondary analyses varying $N$, $P$, and $K$}
\label{sec_sim}
\begin{figure}[!htbp]
\caption{Ninety-five percent variational confidence interval coverage across simulations of different dimensions. Squares are colored according to median coverage (proportion of true values within estimated 95\% variational confidence intervals). Median was taken across all solutions that estimated the correct number of patterns for $K$ = 1, 4, and 10. On the x axis, number of distinct chemicals per pattern is 10 (distinct) or 0 (overlapping). On the y axis, added noise level relative to the true standard deviation increases from 0.2 (low added noise) to 1 (as much added noise as true standard deviation). Each panel describes simulations of a different size, with sample size in columns and mixture size in rows.}
\label{fig:dim_coverage}
\centering
\includegraphics[scale = 0.7]{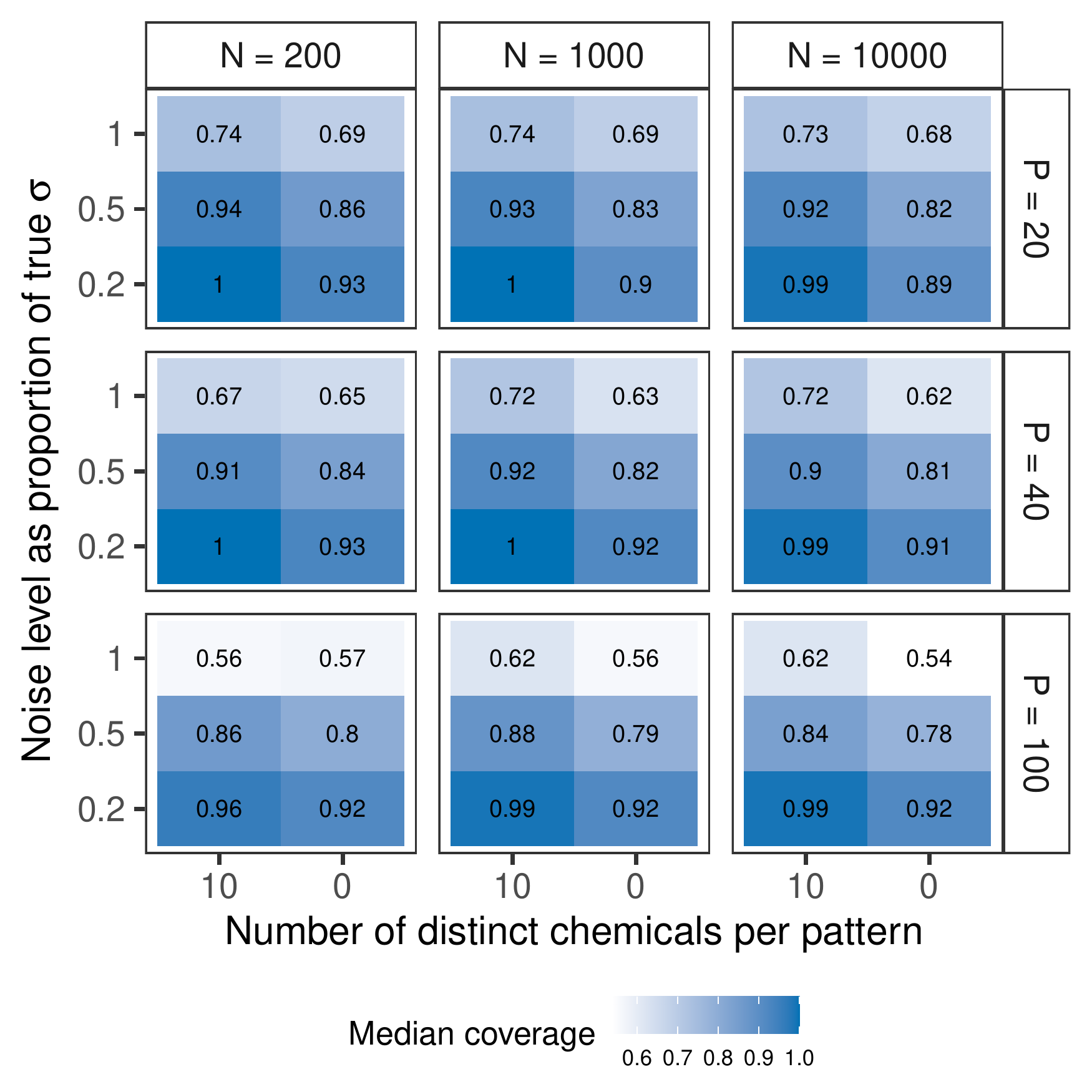}
\end{figure}

These findings largely held in sensitivity analyses with varying dimensions. Simulated datasets with larger sample sizes had higher coverage and lower error, on average, when compared with smaller $N$. Conversely, simulations with larger mixture sizes had lower coverage and higher error, on average, when compare with smaller $P$. Supplemental Figure~\ref{fig:dim_coverage} depicts \bnmf coverage over noise level and separability of patterns across sample and mixture sizes.

The number of patterns affected performance less consistently. \bnmf performed poorly on simulations with one underlying pattern, indicating that \bnmf is not an appropriate method when building a scale or index. PCA, on the other hand, performed much better with only one true pattern. Factor analysis chose the correct number for all simulations with one underlying pattern and chose a ten pattern solution in all but 1.3\% of simulations with ten distinct underlying patterns. On simulations with ten overlapping patterns, performance decreased drastically across all methods. Factor analysis performed the best, estimating the correct number of patterns for 60.8\% of simulations with low noise, but its success rate dropped to 8.9\% for simulations with high noise.

\begin{figure}[!htbp]
\caption{Ninety-five percent variational confidence interval coverage across simulations with different values of $K$. Squares are colored according to median coverage (proportion of true values within estimated 95\% variational confidence intervals). Median was taken across all solutions that estimated the correct number of patterns for all combinations of $N$ = 200, 1,000, and 10,000 and $P$ = 20, 40, and 100. On the x axis, number of distinct chemicals per pattern is 10 (distinct) or 0 (overlapping). On the y axis, added noise level relative to the true standard deviation increases from 0.2 (low added noise) to 1 (as much added noise as true standard deviation). Each panel describes simulations of a different rank, with each column representing a different number of patterns. Gray squares indicate that \bnmf never estimated the correct pattern number.}
\label{fig:pattern_coverage}
\centering
\includegraphics[scale = 0.7]{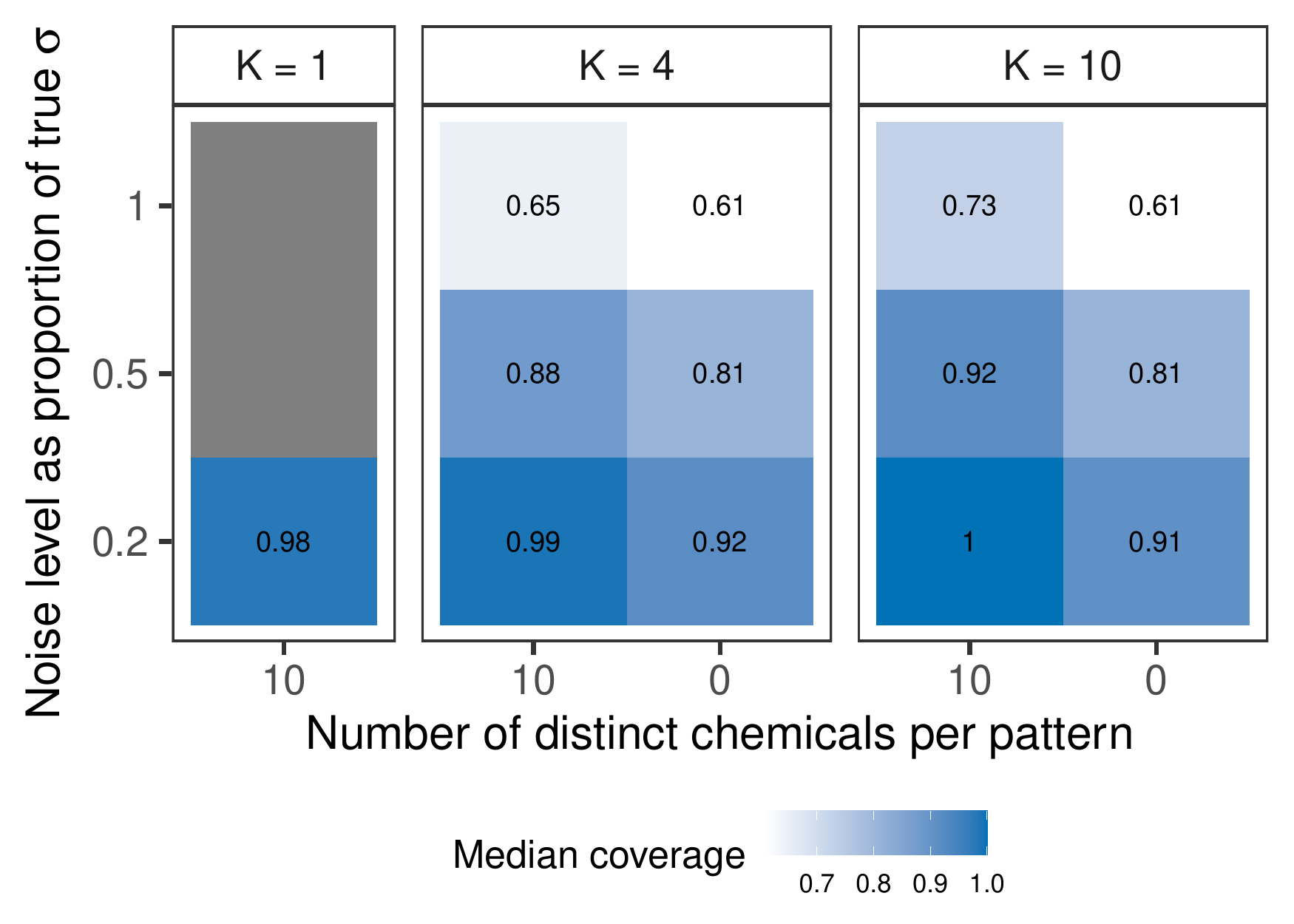}
\end{figure}

Accurate estimation of the underlying number of patterns in these secondary simulations is quantified in Supplemental Table~\ref{tab_rank} as the proportion of simulations for which \bnmfc, NMF, factor analysis, and PCA estimated the correct pattern number. Supplemental Figure~\ref{fig:pattern_coverage} depicts \bnmf coverage over noise level and separability of patterns across values of $K$. We present relative error for predicted values and estimated scores across sample and mixture sizes for $K$ = 1, 4, and 10 in Supplemental Tables~\ref{table:sup1}, \ref{table:sup4}, and \ref{table:sup10}, respectively.

\hspace{5ex}
\begingroup
\renewcommand{\arraystretch}{1.25}
\begin{table}[!hbp] \centering 
  \caption{Percentage of correct estimation of number of patterns describing underlying matrices in secondary simulations. Percentages were taken across 900 simulated datasets (100 datasets of each possible combination of sample sizes (200, 1,000, and 10,000) and mixture sizes (20, 40, and 100). $^*$ FA = factor analysis; NMF-$\ell_2$ = NMF with $\ell_2$ penalty; NMF-P = NMF with Poisson likelihood.} 
  \label{tab_rank}
  \addtolength{\tabcolsep}{-2pt}
\begin{tabular}{lrrr|rr}
\multicolumn{6}{c}{Correct Pattern Number Estimation} \\
\hline 
\hline  
& \multicolumn{3}{c}{Distinct Patterns} & \multicolumn{2}{c}{Overlapping} \\
\hline
\hline  
Model$^*$ & \multicolumn{1}{c}{Rank 1} & \multicolumn{1}{c}{Rank 4} &  \multicolumn{1}{c}{Rank 10} &  \multicolumn{1}{c}{Rank 4} & \multicolumn{1}{c}{Rank 10} \\ 
\hline
\hline  
& \multicolumn{5}{c}{Simulations + 20\% Noise} \\
\hline
\bnmf        &  1.6 & 100.0 & 66.6 & 82.7 & 11.3 \\ 
FA           &  100.0 & 100.0 & 100.0 & 100.0 & 60.8 \\ 
NMF-$\ell_2$ &  100.0 & 100.0 & 92.6 & 70.3 & 0.0 \\ 
NMF-P        &  97.8 & 100.0 & 94.7 & 76.2 & 0.0 \\ 
PCA          &  100.0 & 69.9 & 20.0 & 0.0 & 0.0 \\ 
\hline  
& \multicolumn{5}{c}{Simulations + 50\% Noise} \\
\hline
\bnmf        & 0.0 & 100.0 & 68.2 & 85.2 & 11.6 \\ 
FA           & 100.0 & 100.0 & 100.0 & 100.0 & 28.3 \\ 
NMF-$\ell_2$ & 95.0 & 100.0 & 93.1 & 73.7 & 0.0 \\ 
NMF-P        & 43.6 & 100.0 & 94.9 & 78.0 & 0.0 \\ 
PCA          & 2.8 & 99.3 & 11.3 & 79.3 & 11.8 \\ 
\hline  
& \multicolumn{5}{c}{Simulations + 100\% Noise} \\
\hline
\bnmf        & 0.0 & 87.4 & 81.1 & 90.2 & 16.9 \\ 
FA           & 100.0 & 100.0 & 96.3 & 95.7 & 8.9 \\ 
NMF-$\ell_2$ & 30.7 & 86.3 & 85.3 & 81.7 & 0.4 \\ 
NMF-P        &  3.6 & 93.1 & 87.0 & 83.1 & 1.1 \\ 
PCA          & 0.0 & 0.0 & 0.0 & 0.0 & 0.0 \\ 
\hline \\[-1.8ex] 
\end{tabular} 
\end{table} 
\endgroup

\clearpage
\begingroup
\renewcommand{\arraystretch}{1.35}
\begin{table}[!htbp] \centering 
  \caption{Relative error for predicted values and individual scores of rank 1 secondary simulation with + 20\% noise. Values presented are mean (standard deviation). Error was averaged over 100 simulations for each combination of sample size ($N$ = 200, 10,000), mixture size ($P$ = 20, 40, 100) and simulation strategy (distinct or overlapping patterns). Entries marked as ``---'' indicate that no models on that group of simulations correctly identified the number of patterns, and error on scores could not be computed. $^*$ FA = factor analysis; NMF-$\ell_2$ = NMF with $\ell_2$ penalty; NMF-P = NMF with Poisson likelihood. $^+$ $N$ = sample size.}
  \label{table:sup1} 
 \addtolength{\tabcolsep}{-2pt}
\begin{tabular}{lrr|rr}
\multicolumn{5}{c}{Relative Error for Rank 1 Secondary Simulations with + 20\% Noise} \\
\hline 
\hline  
& \multicolumn{2}{c}{Predicted Values} & \multicolumn{2}{c}{Estimated Individual Scores} \\
\hline
\hline  
Model$^*$ & \multicolumn{1}{c}{$N^+$ = 200} & \multicolumn{1}{c}{$N$ = 10,000} & \multicolumn{1}{c}{$N$ = 200} & \multicolumn{1}{c}{$N$ = 10,000} \\
\hline
\hline  
& \multicolumn{4}{c}{Mixture size $P$ = 20} \\
\hline
BN2MF & 0.21 (0.01) & 0.22 ($<$0.01) & --- & --- \\ 
FA & 0.44 (0.09) & 0.49 (0.02) & 0.75 (0.02) & 0.75 ($<$0.01) \\ 
NMFL2 & 0.17 (0.01) & 0.17 ($<$0.01) & 1.00 ($<$0.01) & 1.00 ($<$0.01) \\ 
NMFP & 0.27 (0.03) & 0.30 (0.01) & 3.07 (0.85) & 11.46 (3.22) \\ 
PCA & 0.23 (0.01) & 0.23 ($<$0.01) & 1.84 (0.07) & --- \\ 
\hline 
& \multicolumn{4}{c}{Mixture size $P$ = 40} \\
\hline 
BN2MF & 0.19 (0.01) & 0.20 ($<$0.01) & --- & --- \\ 
FA & 0.44 (0.09) & 0.49 (0.02) & 0.75 (0.02) & 0.75 ($<$0.01) \\ 
NMFL2 & 0.15 (0.02) & 0.15 ($<$0.01) & 1.00 ($<$0.01) & 1.00 ($<$0.01) \\ 
NMFP & 0.23 (0.03) & 0.21 (0.04) & 2.80 (0.65) & 9.30 (1.92) \\ 
PCA & 0.22 (0.01) & 0.22 ($<$0.01) & 3.01 (0.08) & --- \\ 
\hline 
& \multicolumn{4}{c}{Mixture size $P$ = 100} \\
\hline 
BN2MF & 0.18 (0.01) & 0.18 ($<$0.01) & --- & --- \\ 
FA & 0.44 (0.09) & 0.49 (0.02) & 0.75 (0.02) & 0.74 ($<$0.01) \\ 
NMFL2 & 0.14 (0.02) & 0.14 (0.01) & 1.00 ($<$0.01) & 1.00 ($<$0.01) \\ 
NMFP & 0.19 (0.02) & 0.18 ($<$0.01) & 2.49 (0.55) & --- \\ 
PCA & 0.22 (0.01) & 0.22 ($<$0.01) & --- & --- \\ 
\hline
\hline  
\end{tabular}
\end{table}
\endgroup

\clearpage
\begin{landscape}
\begingroup
\renewcommand{\arraystretch}{1.35}
\begin{table}[!htbp] \centering 
  \caption{Relative error for predicted values and estimated individual scores of rank 4 secondary simulation with + 20\% noise. Values presented are mean (standard deviation). Error was averaged over 100 simulations for each combination of sample size ($N$ = 200, 10,000), mixture size ($P$ = 20, 40, 100) and simulation strategy (distinct or overlapping patterns). $^*$ FA = factor analysis; NMF-$\ell_2$ = NMF with $\ell_2$ penalty; NMF-P = NMF with Poisson likelihood. $^+$ $N$ = sample size.}
  \label{table:sup4} 
 \addtolength{\tabcolsep}{-2pt}
\begin{tabular}{lrr|rr|rr|rr}
\multicolumn{9}{c}{Relative Error for Rank 4 Secondary Simulations with + 20\% Noise} \\
\hline 
\hline  
& \multicolumn{4}{c}{Predicted Values} & \multicolumn{4}{c}{Estimated Individual Scores} \\
\hline
\hline
& \multicolumn{2}{c}{Distinct Patterns} & \multicolumn{2}{c}{Overlapping Patterns} & \multicolumn{2}{c}{Distinct Patterns} & \multicolumn{2}{c}{Overlapping Patterns} \\
\hline
\hline  
Model$^*$ & \multicolumn{1}{c}{$N^+$ = 200} & \multicolumn{1}{c}{$N$ = 10,000} & \multicolumn{1}{c}{$N$ = 200} & \multicolumn{1}{c}{$N$ = 10,000} & \multicolumn{1}{c}{$N$ = 200} & \multicolumn{1}{c}{$N$ = 10,000} & \multicolumn{1}{c}{$N$ = 200} & \multicolumn{1}{c}{$N$ = 10,000} \\
\hline
\hline  
& \multicolumn{8}{c}{Mixture size $P$ = 20} \\
\hline
BN2MF & 0.18 (0.01) & 0.18 ($<$0.01) & 0.22 (0.04) & 0.18 (0.03) & 0.76 ($<$0.01) & 0.94 (0.05) & 0.80 (0.02) & 0.87 (0.04) \\ 
FA & 0.47 (0.05) & 0.48 (0.01) & 0.34 (0.06) & 0.34 (0.02) & 0.76 (0.01) & 0.75 ($<$0.01) & 0.79 (0.01) & 0.79 (0.01) \\ 
NMFL2 & 0.17 (0.01) & 0.16 ($<$0.01) & 0.20 (0.03) & 0.22 (0.03) & 1.00 ($<$0.01) & 1.00 ($<$0.01) & 1.00 ($<$0.01) & 1.00 ($<$0.01) \\ 
NMFP & 0.17 (0.01) & 0.16 ($<$0.01) & 0.20 (0.03) & 0.22 (0.04) & 6.80 (2.46) & 18.80 (5.89) & 5.10 (1.96) & 16.66 (6.39) \\ 
PCA & 0.18 (0.02) & 0.16 ($<$0.01) & 0.19 (0.03) & 0.17 (0.03) & 1.25 (0.07) & 1.23 (0.06) & 1.15 (0.06) & 1.23 (0.03) \\ 
\hline 
& \multicolumn{8}{c}{Mixture size $P$ = 40} \\
\hline 
BN2MF & 0.14 (0.01) & 0.13 ($<$0.01) & 0.15 (0.04) & 0.12 ($<$0.01) & 0.75 (0.02) & 0.90 (0.04) & 0.77 (0.01) & 0.89 (0.03) \\ 
FA & 0.46 (0.05) & 0.47 (0.01) & 0.32 (0.06) & 0.33 (0.01) & 0.76 (0.01) & 0.75 ($<$0.01) & 0.79 (0.01) & 0.78 ($<$0.01) \\ 
NMFL2 & 0.13 (0.01) & 0.12 ($<$0.01) & 0.12 (0.02) & 0.11 ($<$0.01) & 1.00 ($<$0.01) & 1.00 ($<$0.01) & 1.00 ($<$0.01) & 1.00 ($<$0.01) \\ 
NMFP & 0.14 (0.01) & 0.13 ($<$0.01) & 0.13 (0.02) & 0.11 ($<$0.01) & 7.03 (2.52) & 19.16 (5.95) & 5.36 (1.94) & 15.61 (5.02) \\ 
PCA & 0.14 (0.01) & 0.12 ($<$0.01) & 0.13 (0.03) & 0.11 (0.01) & 1.86 (0.08) & 1.85 (0.06) & 1.55 (0.07) & 1.66 (0.04) \\ 
\hline 
& \multicolumn{8}{c}{Mixture size $P$ = 100} \\
\hline 
BN2MF & 0.11 (0.01) & 0.10 ($<$0.01) & 0.09 (0.01) & 0.08 ($<$0.01) & 0.75 (0.02) & 0.95 (0.01) & 0.75 (0.01) & 0.91 (0.01) \\ 
FA & 0.45 (0.06) & 0.46 (0.01) & 0.32 (0.06) & 0.32 (0.01) & 0.75 (0.01) & 0.74 ($<$0.01) & 0.78 (0.01) & 0.78 ($<$0.01) \\ 
NMFL2 & 0.10 (0.01) & 0.09 ($<$0.01) & 0.09 (0.01) & 0.07 ($<$0.01) & 1.00 ($<$0.01) & 1.00 ($<$0.01) & 1.00 ($<$0.01) & 1.00 ($<$0.01) \\ 
NMFP & 0.11 (0.01) & 0.09 ($<$0.01) & 0.09 (0.01) & 0.08 ($<$0.01) & 7.21 (2.55) & 19.36 (5.95) & 5.67 (2.19) & 16.21 (5.26) \\ 
PCA & 0.11 (0.02) & 0.09 ($<$0.01) & 0.09 (0.02) & 0.07 ($<$0.01) & 3.18 (0.11) & 3.20 (0.06) & 2.51 (0.09) & 2.65 (0.03) \\ 
\hline
\hline  
\end{tabular}
\end{table}
\endgroup
\end{landscape}

\clearpage
\begin{landscape}
\begingroup
\renewcommand{\arraystretch}{1.35}
\begin{table}[!htbp] \centering 
  \caption{Relative error for predicted values and individual scores of rank 10 secondary simulation with + 20\% noise. Values presented are mean (standard deviation). Error was averaged over 100 simulations for each combination of sample size ($N$ = 200, 10,000), mixture size ($P$ = 20, 40, 100) and simulation strategy (distinct or overlapping patterns). Entries marked as ``---'' indicate that no models on that group of simulations correctly identified the number of patterns, and error on scores could not be computed. $^*$ FA = factor analysis; NMF-$\ell_2$ = NMF with $\ell_2$ penalty; NMF-P = NMF with Poisson likelihood. $^+$ $N$ = sample size.} 
  \label{table:sup10} 
 \addtolength{\tabcolsep}{-2pt}
\begin{tabular}{lrr|rr|rr|rr}
\multicolumn{9}{c}{Relative Error for Rank 10 Secondary Simulations with + 20\% Noise} \\
\hline 
\hline  
& \multicolumn{4}{c}{Predicted Values} & \multicolumn{4}{c}{Estimated Individual Scores} \\
\hline
\hline
& \multicolumn{2}{c}{Distinct Patterns} & \multicolumn{2}{c}{Overlapping Patterns} & \multicolumn{2}{c}{Distinct Patterns} & \multicolumn{2}{c}{Overlapping Patterns} \\
\hline
\hline  
Model$^*$ & \multicolumn{1}{c}{$N^+$ = 200} & \multicolumn{1}{c}{$N$ = 10,000} & \multicolumn{1}{c}{$N$ = 200} & \multicolumn{1}{c}{$N$ = 10,000} & \multicolumn{1}{c}{$N$ = 200} & \multicolumn{1}{c}{$N$ = 10,000} & \multicolumn{1}{c}{$N$ = 200} & \multicolumn{1}{c}{$N$ = 10,000} \\
\hline
\hline  
& \multicolumn{8}{c}{Mixture size $P$ = 20} \\
\hline
BN2MF & 0.42 (0.03) & 0.29 (0.03) & 0.41 (0.03) & 0.37 (0.02) & --- & 0.93 (0.02) & --- & --- \\ 
FA & 0.51 (0.04) & 0.51 (0.01) & 0.40 (0.05) & 0.39 (0.01) & 0.78 (0.01) & 0.77 ($<$0.01) & --- & --- \\ 
NMFL2 & 0.27 (0.02) & 0.25 ($<$0.01) & 0.23 (0.02) & 0.22 ($<$0.01) & 1.00 ($<$0.01) & 1.00 ($<$0.01) & --- & --- \\ 
NMFP & 0.27 (0.02) & 0.25 ($<$0.01) & 0.25 (0.02) & 0.23 (0.01) & 8.76 (2.53) & 24.15 (8.01) & --- & --- \\ 
PCA & 0.35 (0.02) & 0.33 ($<$0.01) & 0.28 (0.02) & 0.27 (0.01) & --- & --- & --- & --- \\ 
\hline 
& \multicolumn{8}{c}{Mixture size $P$ = 40} \\
\hline 
BN2MF & 0.32 (0.05) & 0.20 ($<$0.01) & 0.35 (0.03) & 0.29 (0.02) & 0.90 ($<$0.01) & 0.96 (0.03) & --- & --- \\ 
FA & 0.48 (0.04) & 0.48 (0.01) & 0.37 (0.05) & 0.37 (0.01) & 0.77 (0.01) & 0.76 ($<$0.01) & --- & 0.81 (0.01) \\ 
NMFL2 & 0.19 (0.01) & 0.18 ($<$0.01) & 0.17 (0.01) & 0.16 ($<$0.01) & 1.00 ($<$0.01) & 1.00 ($<$0.01) & --- & --- \\ 
NMFP & 0.19 (0.01) & 0.18 ($<$0.01) & 0.18 (0.01) & 0.17 ($<$0.01) & 9.16 (2.69) & 24.19 (8.01) & --- & --- \\ 
PCA & 0.28 (0.02) & 0.27 (0.05) & 0.23 (0.02) & 0.21 (0.01) & --- & 1.26 (0.05) & --- & --- \\ 
\hline 
& \multicolumn{8}{c}{Mixture size $P$ = 100} \\
\hline 
BN2MF & 0.18 (0.06) & 0.13 ($<$0.01) & 0.27 (0.03) & 0.14 (0.03) & 0.89 ($<$0.01) & 0.92 (0.01) & --- & 0.93 (0.01) \\ 
FA & 0.47 (0.04) & 0.47 (0.01) & 0.36 (0.05) & 0.36 (0.01) & 0.76 (0.01) & 0.75 ($<$0.01) & --- & 0.80 (0.01) \\ 
NMFL2 & 0.14 ($<$0.01) & 0.13 ($<$0.01) & 0.13 (0.01) & 0.12 ($<$0.01) & 1.00 ($<$0.01) & 1.00 ($<$0.01) & --- & --- \\ 
NMFP & 0.15 (0.01) & 0.13 ($<$0.01) & 0.14 (0.01) & 0.12 ($<$0.01) & 9.87 (3.05) & 25.12 (8.24) & --- & --- \\ 
PCA & 0.22 (0.02) & 0.13 ($<$0.01) & 0.18 (0.02) & 0.17 (0.01) & 1.99 (0.03) & 1.97 (0.05) & --- & --- \\ 
\hline
\hline  
\end{tabular}
\end{table}
\endgroup
\end{landscape}

\clearpage
\subsection{\bnmf variational confidence interval coverage for secondary simulations}

\end{document}